%% file: main.tex
\documentclass[acmsmall, nonacm]{acmart}
\usepackage{graphicx} 
\usepackage{url}
\usepackage{graphicx}
\usepackage{textcomp}
\usepackage{xcolor}
\def\BibTeX{{\rm B\kern-.05em{\sc i\kern-.025em b}\kern-.08em
    T\kern-.1667em\lower.7ex\hbox{E}\kern-.125emX}}
\usepackage{natbib}
\setcitestyle{square, comma, numbers,sort&compress, super}
\usepackage[T1]{fontenc}
\usepackage{listings}
\AtBeginDocument{\DeclareCaptionSubType{lstlisting}}
\usepackage{caption,subcaption}
\usepackage[binary-units=true]{siunitx} 
\usepackage[frozencache]{minted}
\usepackage{listings}
\setminted{fontsize=\small,baselinestretch=1}
\usepackage{comment}
\usepackage[printonlyused, withpage]{acronym}
\usepackage{booktabs}

\usepackage[ruled, lined, linesnumbered, commentsnumbered, longend]{algorithm2e}
\usepackage{nicematrix}

\usepackage{ifthen}

\usepackage{newtxtext,newtxmath}
\usepackage{xcolor}
\usepackage{amsmath,amssymb,amsfonts}

\renewcommand\footnotetextcopyrightpermission[1]{} 
\begin{document}

\title[CNM/CIM Landscape]{The Landscape of Compute-near-memory and Compute-in-memory: A Research and Commercial Overview}
\author{Asif Ali Khan}
\affiliation{%
  \institution{TU Dresden}
  \city{Dresden}
  \country{Germany}}
\email{asif_ali.khan@tu-dresden.de}
\orcid{0000-0002-5130-9855}

\author{João Paulo C. de Lima}
\affiliation{%
  \institution{TU Dresden and ScaDS.AI}
  \city{Dresden}
  \country{Germany}}
\email{joao.lima@tu-dresden.de}
\orcid{0000-0001-9295-3519}

\author{Hamid Farzaneh}
\affiliation{%
  \institution{TU Dresden}
  \city{Dresden}
  \country{Germany}}
\email{hamid.farzaneh@tu-dresden.de}
\orcid{0000-0002-1780-6217}

\author{Jeronimo Castrillon}
\affiliation{%
  \institution{TU Dresden and ScaDS.AI}
  \city{Dresden}
  \country{Germany}}
\email{jeronimo.castrillon@tu-dresden.de}
\orcid{0000-0002-5007-445X}

\date{October 2023}
\begin{abstract}
In today's data-centric world, where data fuels numerous application domains, with machine learning at the forefront, handling the enormous volume of data efficiently in terms of time and energy presents a formidable challenge. Conventional computing systems and accelerators are continually being pushed to their limits to stay competitive. In this context, computing near-memory (CNM) and computing-in-memory (CIM) have emerged as potentially game-changing paradigms. 
This survey introduces the basics of CNM and CIM architectures, including their underlying technologies and working principles. We focus particularly on CIM and CNM architectures that have either been prototyped or commercialized.
While surveying the evolving CIM and CNM landscape in academia and industry, we discuss the potential benefits in terms of performance, energy, and cost, 
along with the challenges associated with these cutting-edge computing paradigms.
\end{abstract}

\maketitle

\section{Introduction}
\label{sec:intro}
\input{contents/01-intro}

\section{Terminology and background}
\label{sec:terminology}
\input{contents/02-background}

\section{Technology overview} 
\label{sec:tech-overview}
\input{contents/tech-overview}

\section{Selected architectures} 
\label{sec:select-arch}
\input{contents/selected-arch}

\section{Commercial landscape} 
\label{sec:commercial}
\input{contents/cim-startups}

\section{Conclusions} 
\label{sec:conclusions}
\input{contents/conclusions}

\section*{Acknowledgements} 
This work was supported by Vsquared Ventures (VSQ).
Special thanks to Max Odendahl (Venture Partner at VSQ) for his feedback on previous versions of the manuscript.
This work was also supported by the German Research Council (DFG) through the HetCIM project (project number 502388442) in the context of the DFG Priority Program on Disruptive Memory Technologies (SPP2377 \url{https://spp2377.uos.de}) and the German Federal Ministry of Education and Research (BMBF, project number 01IS18026A-D) by funding the competence center for Big Data and AI ScaDS.AI Dresden/Leipzig (\url{https://scads.ai}). 
\bibliographystyle{IEEEtran}
\bibliography{main}

\input{acronyms}

\end{document}

%% file: contents/01-intro.tex
In conventional computing systems, the processor and memory are two independent entities connected via communication pathways, known as buses. 
When the CPU processes data, it requires fetching it from memory via the bus, conducting the necessary computations, and subsequently storing the results back in memory. This off-chip communication becomes a limiting factor for data-intensive workloads due to the limited transfer rate and high energy per bit of buses.  
For example, the data transfer between the logic (CPUs and GPUs) and memory chips (DRAM or flash memory) requires approximately 10–100 times more energy than the logic operation itself~\cite{cim_motiv}. \emph{Compute-near-memory} (CNM) and \ac{cim} concepts address this bottleneck by enabling computations close to where the data resides. This is achieved either by implementing CMOS logic on or closer to the memory chip, or by leveraging the inherent physical properties of memory devices to perform computations in place. 

The core concept behind \acs{cnm}/\ac{cim} is not entirely new. However, the sudden surge in these systems can be attributed to two primary factors, namely, 
the exponential increase in the volume of data required for modern applications, and 
the technological readiness. Recent advancements in machine learning, particularly the emergence of generative AI and \ac{llm}, demand the processing of terabytes of data, substantial computational resources, and complex execution, thus highlighting the limitations of traditional computing systems. A recent study revealed that OpenAI utilized over 3600 of NVIDIA's HGX A100 servers, totaling around 29,000 GPUs, to train ChatGPT, resulting in a daily energy consumption of 564 MWh~\cite{ai-energy-footprint}. 
Projections indicate that by 2027, AI is expected to consume between 85 and 124 TWh annually, equivalent to approximately 0.5\% of the world's total electricity consumption. It is no surprise that Microsoft has announced plans to develop its own nuclear reactors to power their data centers~\cite{verge_article}. 

Currently, machine learning applications primarily leverage GPU accelerators like A100, H100, GH200, application-specific integrated circuits (e.g., Google's TPU), and dataflow processors as in the case of companies like GraphCore, Cerebras, Groq, and SambaNova~\cite{ai-accel-review-22}.
Over the past few years, \acs{cnm}/\ac{cim} systems have also transcended their prototypical stages and successfully entered the market. 
The timing of these advancements in CIM/CNM systems is of paramount importance as it perfectly aligns with the AI revolution. As a result, numerous companies have emerged in the last few years offering CIM/CNM solutions for various use domains. 
This surge reflects a competitive landscape where these companies are striving to leverage the demand and cater to various market segments. All commercially available solutions hold the promise of significantly reducing execution time and energy consumption for data-intensive workloads.

This survey explores CNM and CIM architectures, detailing their technologies, fundamental concepts, working principles, and the evolving landscape in academia and industry. Addressing a broader audience, it provides foundational concepts for non-experts while delivering state-of-the-art insights for experts in the domain. It also summarizes the impact and challenges associated with adopting the novel CIM/CNM computing paradigms. Concretely, our discussion revolves around three key aspects:

\begin{enumerate}
     \item \textbf{Key technologies and concepts:} In CNM systems, a specialized CMOS logic is integrated into the memory chip. This logic can be either general-purpose, as in UPMEM systems~\cite{upmem}, or domain-specific, as in systems developed by Samsung~\cite{samsung-ISSCC,samsung-ISCA} and SK Hynix~\cite{aim,aim-ISSCC}, integrated within DRAM memory chips. 
     While CNM significantly reduces data movement, it does not eliminate it.
    In contrast, CIM nearly eliminates data movement by performing computations within the same devices that store the data. 
    A particularly noteworthy operation is the analog dot-product in memory, which is of significant importance to the machine learning domain and can be performed in constant time. Initially demonstrated in crossbar-configured resistive \ac{nvm}  technologies like \ac{pcm}~\cite{pcm-prog1} and \ac{rram}~\cite{ReRAM}, this concept has also been shown with SRAM, \ac{mram}~\cite{MRAM}, and \ac{fefet}~\cite{FeFET}.
     While other arithmetic, search and boolean logic operations have also been demonstrated using CIM, they have received comparatively less attention.
    
    \item \textbf{Commercial trends:} As the demand for fast and efficient computing systems continues to rise, the in-/near-memory computing market is experiencing rapid expansion. In 2022, this market was valued at USD 15.5 billion, with an anticipated \ac{cagr} of 17.5\% over the next decade~\cite{market-study}. This growth is underscored by the proliferation of startups offering CIM and CNM solutions. Some of these companies have secured hundreds of millions of dollars in early funding rounds. While many of these companies provide innovative solutions for data-intensive applications (dominated by AI inference), there is no clear winner yet.
    At present, these solutions are predominantly based on SRAM technology, although solutions based on resistive NVM and flash technologies also exist~\cite{PIM-tipping-point}. This trend can be attributed to the mature tools and design processes for SRAM compared to emerging NVMs. However, considering the SRAM's scalability aspects and its static power consumption, it is likely that NVMs, particularly \ac{pcm}, \ac{rram}, \ac{mram}, and \ac{fefet}, will progressively replace or complement SRAM as these technologies mature. 
    
     \item \textbf{Challenges: } Although CIM/CNM systems are at the tipping point, they are yet to make substantial inroads into the market. The predominant obstacle facing these systems is perhaps the absence of a software ecosystem, which renders programmability and optimization exceedingly challenging. This is also highlighted by a recent Meta article~\cite{facebook-accel} stating, \emph{We’ve investigated applying processing-in-memory (PIM) to our workloads and determined there are several challenges to using these approaches. Perhaps the biggest challenge of PIM is its programmability}.
     Other challenges requiring attention include: addressing reliability concerns associated with emerging NVMs (particularly in the CIM context), developing novel performance models, profiling and analysis tools for these systems, which could be leveraged to exploit their potential effectively.
 \end{enumerate}

The remainder of this paper is structured as follows: Section~\ref{sec:terminology} explains the terminology associated with these domains and provides insights into the conventional Von-Neumann computing approach, as well as the emerging data-centric paradigms. In Section~\ref{sec:tech-overview}, a comprehensive overview of promising memory technologies within the context of \ac{cim} and \acs{cnm} systems is provided. Section~\ref{sec:select-arch} outlines various common \ac{cim} and \acs{cnm} systems including very recent prototype chips from various industries. Lastly, Section~\ref{sec:commercial} presents a comprehensive overview of the commercial landscape for these systems (start-ups), discussing their products details, target application domain, and funding status. Finally, Section~\ref{sec:conclusions} concludes the paper by summarizing our key observations and providing insights and recommendations into the future.

%% file: contents/02-background.tex
This section highlights the bottleneck in the Von Neumann computing model by discussing its working mechanism, motivates the need for memory-centric computing, and explains the terminology. 

\subsection{Mainstream Von-Neumann Computing}
\label{subsec:vonNeumann}
\begin{figure}[tbh]
\centering
\includegraphics[scale=0.11]{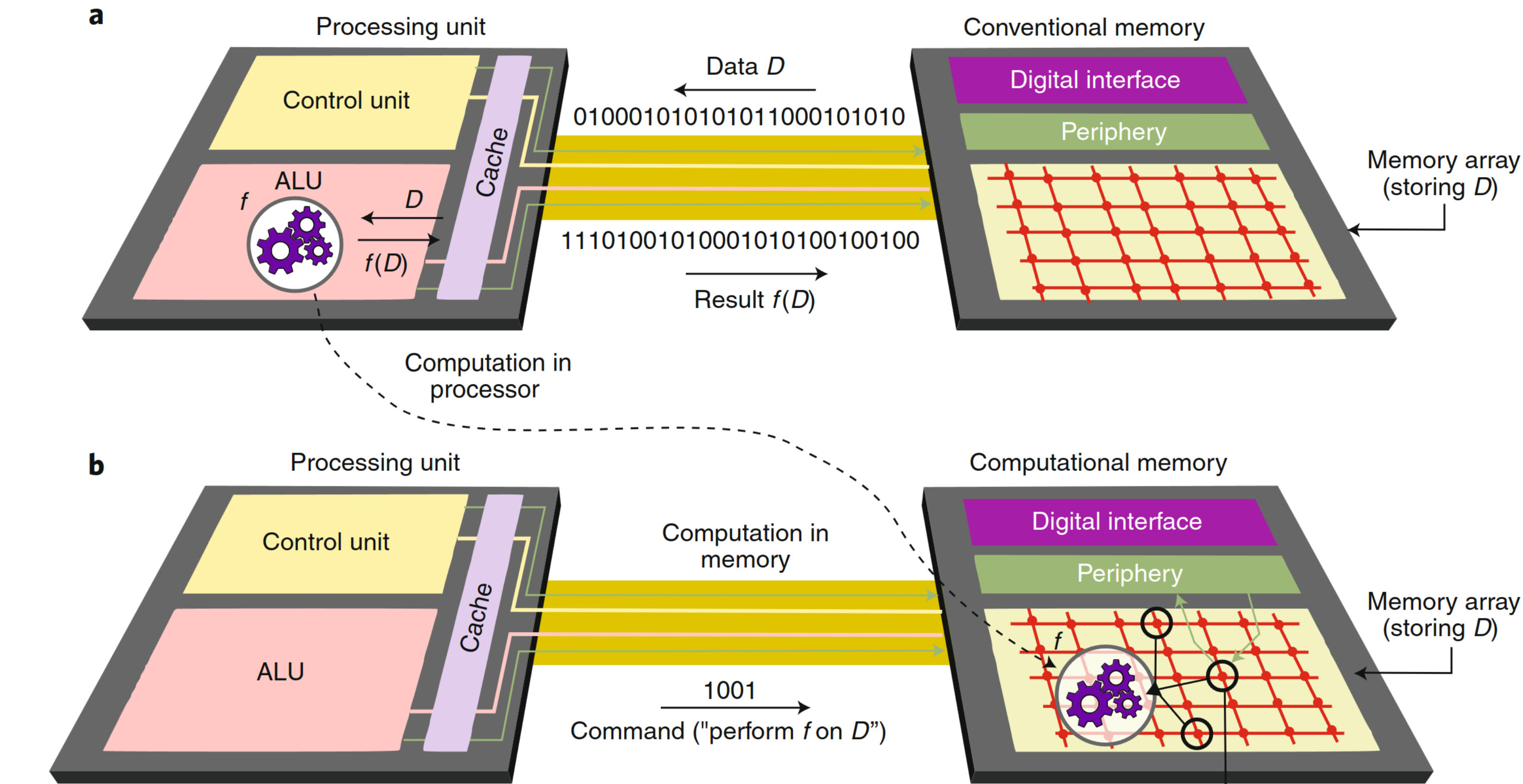}
\caption{(a) Conventional computing system where an operation $f$ is performed on data $D$ in the CPU (b) Memory-centric design where $f$ is computed in the proximity of $D$ and CPU is mainly working as a control unit~\cite{cim_survey}.} 
\label{fig:mem-centric}
\end{figure}

As depicted in Figure~\ref{fig:mem-centric}a, the interaction between memory and the processor in the Von Neumann architecture is facilitated through address and data buses. 
However, because CPU performance significantly outpaces memory performance, the Von Neumann models are often bottlenecked by the memory. To address this challenge and mitigate the impact of larger memory access latencies on the CPU, modern processors incorporate a tiered hierarchy of caches. Caches are smaller memory units that, while being much smaller compared to main memory and storage, are notably faster. The first-level cache (L1) is typically integrated onto the CPU chip and operates nearly at CPU speed, enabling single-cycle access. L2 cache is usually shared by multiple cores and can vary in location depending on the design goals. Some systems even include an L3 cache, usually larger and situated off-chip.

Note that Von Neumann architectures are characterized by the sequential execution of instructions. However, in multi-core CPU systems, parallel execution of instructions at various levels of granularity, including instruction-level, data-level, and thread-level parallelisms, is supported. To enhance performance and energy efficiency in resource-constrained systems, specialized accelerators are often developed and integrated on the same chip. For instance, in application domains such as embedded systems, digital signal processing, and networking, \ac{mpsoc} architectures are employed. These integrate multiple processor cores, memory, input/output interfaces, and potentially specialized hardware accelerators, enabling parallel execution of different tasks to meet specific constraints. 

Although these designs may significantly enhance performance when compared to conventional CPU-only systems, the underlying design principle remains CPU-centric and follows the Von Neumann model of execution. Consequently, the performance improvements, largely resulting from concurrent execution, heavily rely on the nature of the application. In cases where an application is memory-bound, i.e., most of the execution time is spent on the memory accesses and not on the actual compute operations, the shared data bus is fully occupied and becomes a bottleneck. Even for compute-bound applications, where these architectures can yield substantial gains in execution time, power consumption remains largely unaffected and might even increase due to the complex structure of these systems.

\subsection{Memory-centric computing}
\label{subsec:mem-centric}
Unlike conventional computing systems where CPU has a central role and is responsible for all computations, most computations in the memory-centric designs are performed within or near memory. 
As depicted in Figure~\ref{fig:mem-centric}b, the core concept revolves around minimizing data transfer on the bus by relocating a substantial share of computations closer to the data (memory). The CPU's primary role becomes issuing commands and handling computations that cannot be effectively executed in close proximity to the memory.

The concept of memory-centric computing is not a novel one, but it has experienced a significant surge in recent years. Consequently, various terms have emerged, often referring to the same idea, and more detailed classifications have been introduced in architectural designs. 
This section aims to clarify the terminology surrounding these approaches.

\subsubsection{Compute-in-memory}
Computing systems can be broadly divided into two categories: \acf{cim} systems and \ac{com} systems (see Section~\ref{subsec:vonNeumann}).
In the literature, there are different names for similar things. These architectures are often named based on (1) the \emph{location} of the compute units within the memory hierarchy (near cache or near main memory), or (2) based on the underlying paradigm, i.e., whether the memory device itself is used to implement computation (\ac{cim}), or whether extra CMOS-logic is added near the memory to perform computations (\ac{cnm}). 
Figure~\ref{fig:cim-cnm-com} shows an overview of different processor and memory system configurations. 
A compute operation can be a logic operation or an arithmetic operation such as addition and multiplication.
\ac{cim} systems are also frequently referred to as \ac{imc}, \ac{imp}, \ac{pim}, \ac{pum} or \ac{lim} systems~\cite{LiM}. For the purposes of this report, we will use the term \ac{cim}. 

\subsubsection{Compute-near-memory}
\begin{figure}[tbh]
\centering
\includegraphics[scale=0.2]{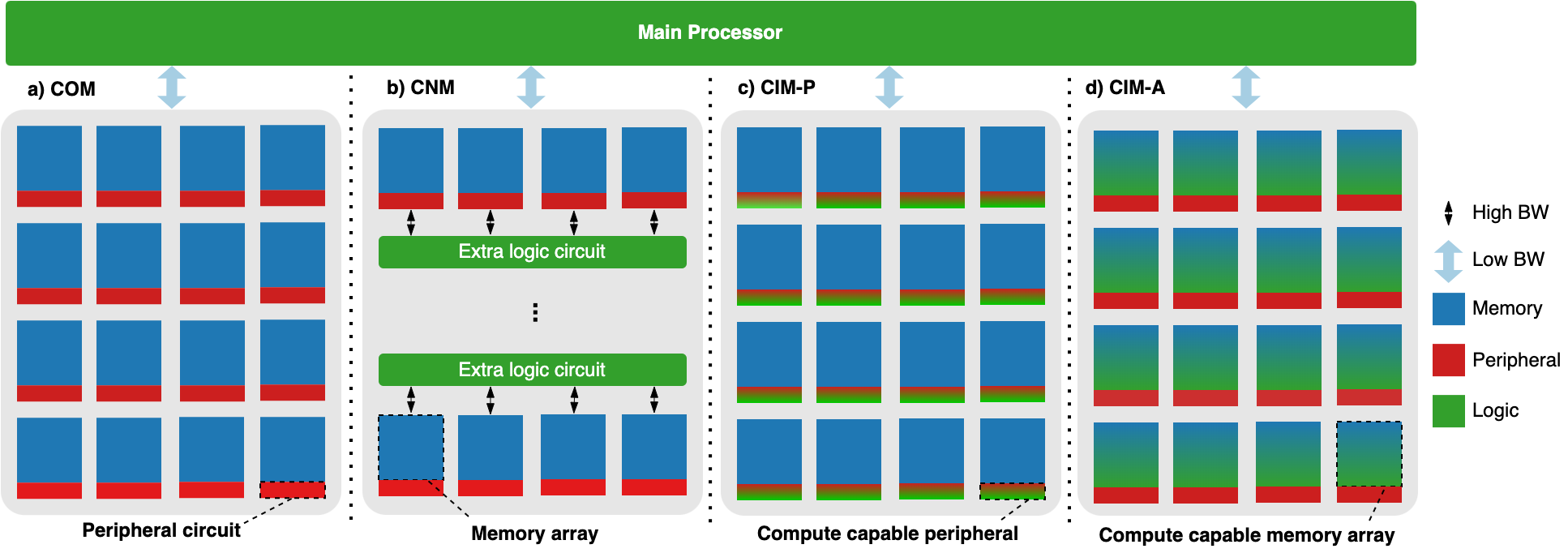}

\caption{High-level overview of systems where computation is performed a) COM (mainstream computing): outside of memory system, b) \ac{cnm}: using a logic connected to the memory via the memory high-bandwidth channel, c) \ac{cim-p}: in the memory peripheral circuitry, and d) \ac{cim-a}: using memory cells within the memory array.} 
\label{fig:cim-cnm-com}
\end{figure}

 When operations are computed outside the memory (\ac{com}) using conventional computing cores (Fig~\ref{fig:cim-cnm-com}.a), the architecture is a conventional Von Neumann system (see Section~\ref{subsec:vonNeumann}). On the other hand, if the computations are performed outside the memory but with a dedicated logic unit connected to the memory module via a high-bandwidth channel (Fig~\ref{fig:cim-cnm-com}.b), the system is referred to as a \acf{cnm} or \ac{nmc} or \ac{nmp}, or \ac{pnm} system. In this report, we will restrict ourselves to calling it \ac{cnm}.

In the \ac{cim} category, computations can be carried out using memory cells within the memory array, known as \ac{cim-a} (see Fig~\ref{fig:cim-cnm-com}.d). Alternatively, computations can occur in the memory peripheral circuitry, termed \ac{cim-p} (see Fig~\ref{fig:cim-cnm-com}.c). 

In \ac{cim-a}, memory cells are often modified to support logic design, e.g., in~\cite{magic}. Sometimes, it also necessitates changes to the periphery to support the modified cells. Therefore, some literature further divides the \ac{cim-a} designs into basic \ac{cim-a} that do not require any modifications to the periphery, e.g.,~\cite{sf-logic}, and hybrid \ac{cim-a} that requires support from the peripheral circuit.  A well-known example of a hybrid \ac{cim-a} is the MAGIC design~\cite{magic} that requires extending the peripheral circuit to write multiple memory rows simultaneously. 

\begin{figure*}[tbh]
\centering
\includegraphics[scale=0.5]{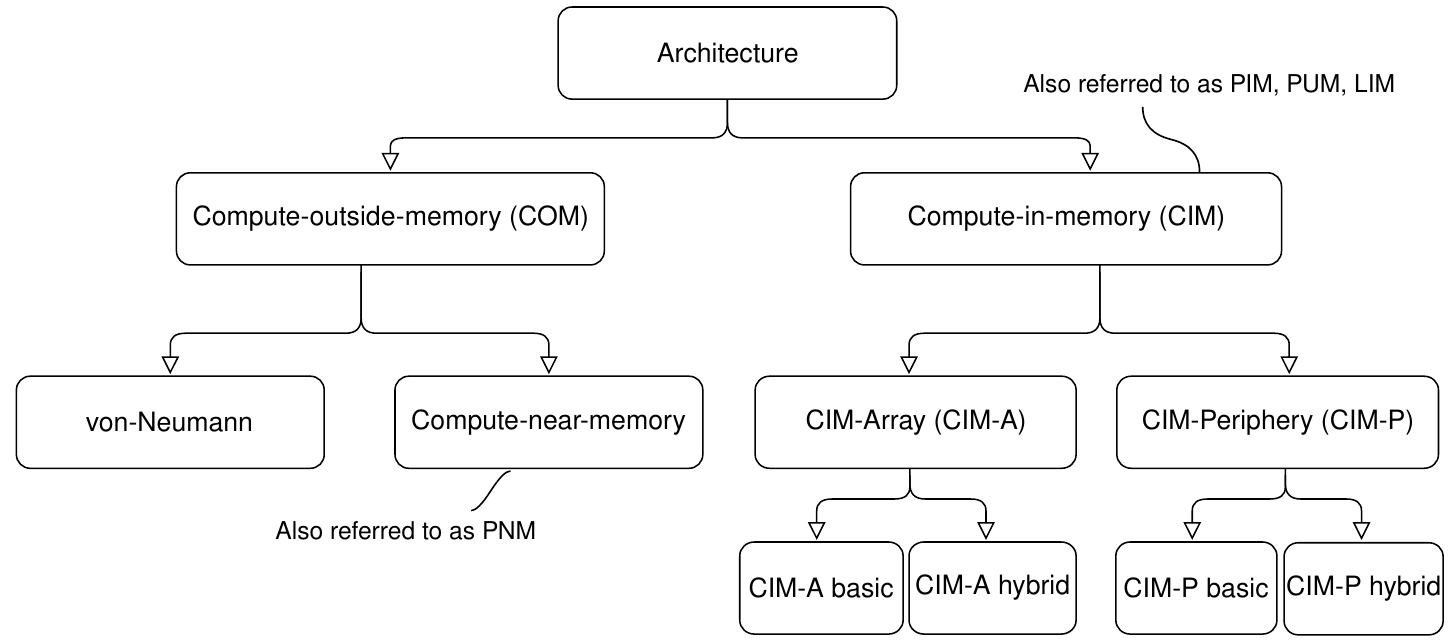}
\caption{\ac{cim} and \ac{cnm} classification.} 
\label{fig:term}
\end{figure*}

Typical examples of \ac{cim-p} architectures are crossbars employing \acp{adc} and \acp{dac} to implement \ac{mvm} and other logic operations~\cite{prime, isaac}. Additionally, \ac{cim-p} designs employing customized sense amplifiers also exist~\cite{li2016pinatubo}. Similar to \ac{cim-a}, \ac{cim-p} can be either basic, as in Pinatubo~\cite{li2016pinatubo}, requiring no changes to the memory array, or hybrid, as seen in ISAAC~\cite{isaac}. A summary of the terminology's classification is presented in Figure~\ref{fig:term}.

Both \ac{cim-a} and \ac{cim-p} can also be used together, wherein the memory array calculates partial results that are later post-processed or accumulated in the peripheral circuit.  In such cases, it is referred to as a \ac{cim} architecture.

%% file: contents/tech-overview.tex
In this section, we present an overview of the main memory cells used in various \ac{cnm} and \ac{cim} systems, encompassing both volatile \ac{sram}, \ac{dram}) and non-volatile types such as \ac{pcm}, \ac{mram}, \ac{rram} and \acp{fefet}. 
These memory technologies are versatile enough to serve as main memory for data storage and support \ac{cnm} without requiring any modification to the memory chips.
Additionally, we explore how these memory cells can be leveraged for in-memory computation, considering technological aspects such as performance, energy consumption, lifetime, CMOS compatibility, and other relevant factors. 
Before going into the individual technologies, let us first explain the different components of the memory subsystem.


\subsection{Memory subsystem}
\label{ss:mem-subsystem}
The memory subsystem typically consists of a four-level hierarchy, each with its own characteristics and access time. The fastest and smallest level is the CPU registers, while the slowest one in the hierarchy is storage devices (HDD/SSD). The \ac{cnm}/\ac{cim}  concepts have been proposed at different levels in the memory hierarchy, e.g., in-cache computing~\cite{in-cache-comp}, in-DRAM~\cite{ambit}, in-storage computing~\cite{in-storage-comp}. However, \ac{cnm}/\ac{cim} at the main memory level have rightly gained more attention than others. 

\begin{figure}[tbh]
\centering
\includegraphics[scale=1.7]{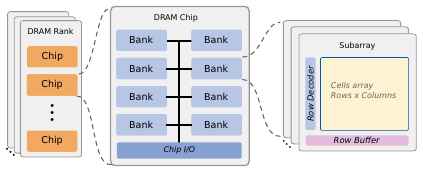}
\caption{Typical \ac{dram} system organization.} 
\label{fig:mem-org}
\end{figure}

The main memory is also typically organized hierarchically, as shown in Figure~\ref{fig:mem-org}. The main components of the hierarchy are memory cells, subarrays, banks, and ranks. 
A cell is the fundamental working unit and serves as a building block to form arrays. An array is a 2D grid of cells connected via \emph{word lines} and \emph{bitlines}. A word line is a horizontal line that connects all the memory cells in a row, while bitlines, on the other hand, connect all cells in a column. Memory cells are placed at the intersection of word lines and bitlines.
The combination of a specific word line and a bit line uniquely identifies each memory cell. When a particular memory cell needs to be accessed, the corresponding word line and bit line are activated, allowing data to be read from or written into that cell.

\subsection{\ac{dram}}
\label{ss:dram}
\Ac{dram} is the most mature and widely used memory technology today. 
A \ac{dram} cell is composed of a transistor and a capacitor. When the capacitor is fully charged, it represents the logical value 1, while a discharged capacitor represents the logical value 0. To access data from the \ac{dram} array, the memory controller brings a particular row into the row buffer by sending an \emph{activate} command. A \emph{read} command is then issued to read specific column(s) from the row buffer and put them on the bus. 

\ac{dram} has scaled nicely for decades and has been used across application domains and systems ranging from HPC to portable devices. However, it is presently facing several challenges to remain the dominant technology. The increasing demand for higher capacity has put tremendous pressure on the \ac{dram} capacitor size to shrink which makes it susceptible to errors. Also, the increase in capacity is significantly increasing the refresh power budget.

To address the escalating demands for higher bandwidth in modern applications, 3D stacked \ac{dram} architectures, such as \ac{hbm} (see Section~\ref{sss:fimdram}), have been proposed. 
These architectures consist of stacked \ac{dram} dies atop a logic layer, interconnected through \ac{tsvs}, resulting in remarkably higher bandwidth. These structures are also employed in a series of \ac{cnm} solutions, where the logic layer is used to implement custom logic and perform computations in closer proximity to the data~\cite{cnm-review-gagan, delft-cim-survey}.

From the \ac{cim} perspective, the majority of in-\ac{dram} implementations rely on charge sharing, wherein multiple rows are activated simultaneously. The shared charge is then utilized in a controlled manner to carry out various logic and data copy operations~\cite{ambit}. Moreover, cleverly manipulating the memory timing parameters, deviating from standard timings, has also been employed to implement different logic operations~\cite{computedram}.



\subsection{SRAM}
\label{ss:sram}
SRAM is another mature memory technology that provides fast and efficient memory accesses. It is commonly used in caches, register files, and other high-speed memory applications where speed and low access latency are critical.
An \ac{sram} cell consists of multiple transistors arranged in a specific configuration to hold one bit of data. 
The most common configuration of an \ac{sram} cell is a pair of cross-coupled inverters that are connected in a feedback loop, forming a latch.

\begin{figure}[tbh]
\begin{center}
\subfloat[A sample 6T \ac{sram} cell~\cite{imac}\label{fig:sram-cell} ]{%
      \includegraphics[scale=0.1]{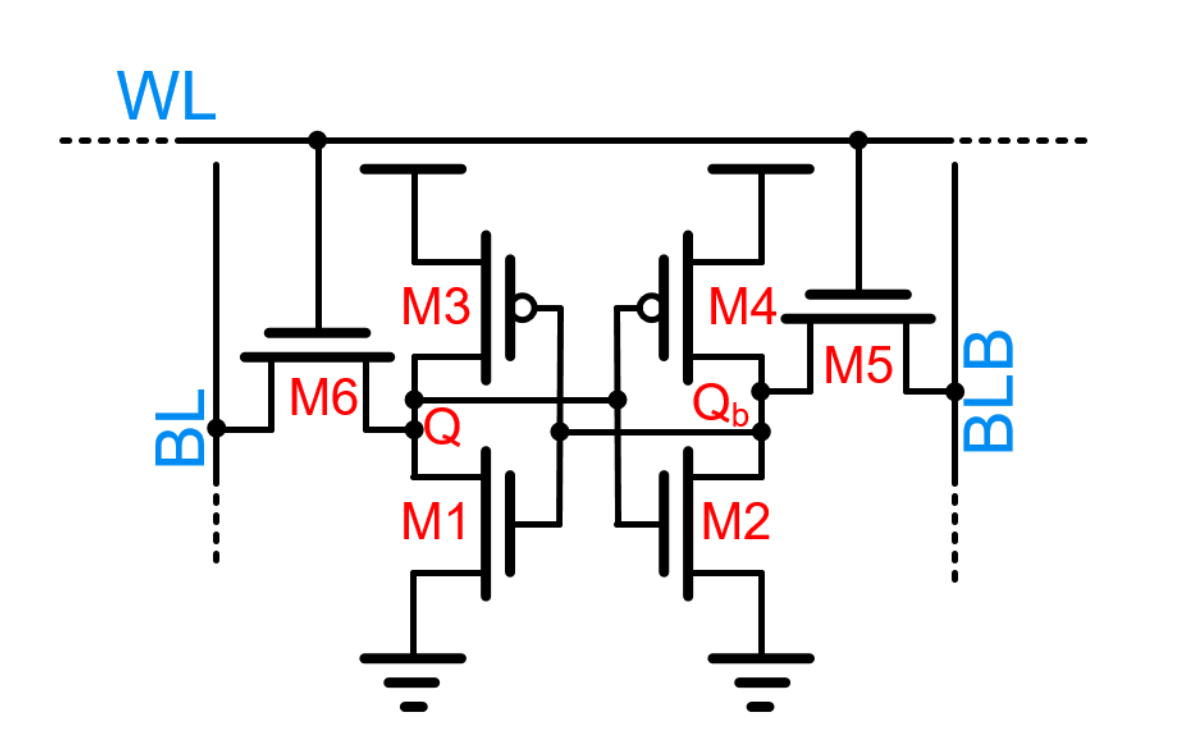}
     }
\subfloat[Metal-oxide RRAM cell~\cite{wang2016functionally} \label{fig:reram}]{%
      \includegraphics[scale=0.2]{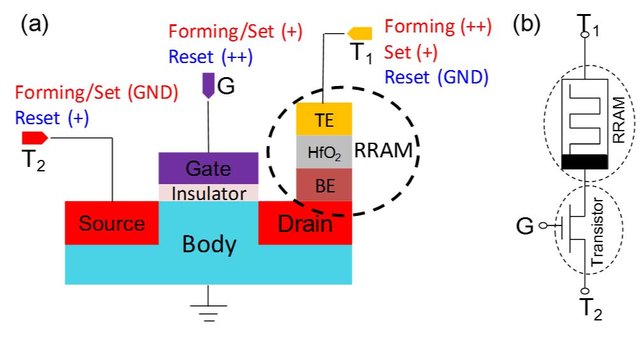}
     }
\subfloat[\ac{fefet} device~\cite{fefet-hdc} \label{fig:fefet}]{%
      \includegraphics[scale=.12]{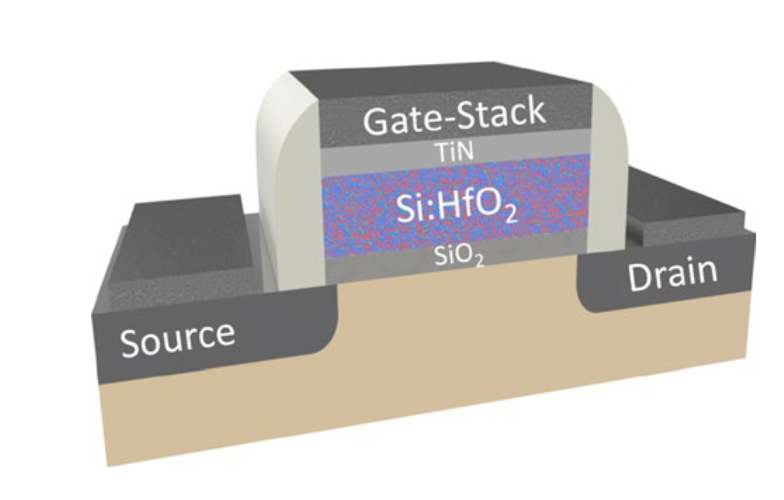}
     }
\end{center}
\caption{Cell structures of various memory technologies}
\label{fig:mem-cells}
\vspace{-.2cm}
\end{figure}

To read and store data on the cell, the bitline terminals, \ac{bl} and \ac{blb} (see Fig.~\ref{fig:sram-cell}), are precharged and discharged, and the wordline(\ac{wl}) is activated or deactivated depending on the values reading/writing from/to the cell. 

There have been proposals for in-\ac{sram} computing, especially at the last-level cache, which can be considerably slower compared to the L1 cache (e.g., by an order of magnitude). Similar to \ac{dram}, most in-\ac{sram} computing architectures also leverage charge sharing in the bitlines. Specifically, precharging the bitlines in a controlled manner and activating multiple rows simultaneously enables performing logic operations~\cite{in-cache-comp}. For bitwise multiplication in \ac{sram}, research has demonstrated that the amplitude of the input voltage at the \ac{wl} directly influences the discharge rate of the \ac{blb}.
The voltage discharge on \ac{blb}, achieved within a specific timeframe, effectively represents a one-bit multiplication of the data stored in the \ac{sram} cell.

\subsection{Phase change memory (\ac{pcm})}
\label{subsec:pcm}
\ac{pcm} is resistive memory technology that employs reversible phase changes in materials to store data. The earliest demonstration of a 256-bit \ac{pcm} prototype dates back to 1970~\cite{pcm}. Today, \ac{pcm} stands as one of the most extensively researched \ac{nvm} technologies. A \ac{pcm} device comprises a phase-changing material sandwiched between two electrodes (very similar to Fig.~\ref{fig:reram}), which transitions between crystalline (low resistance state) and amorphous (high resistance state) phases. These two resistance states represent binary logic states, i.e., 1 and 0.

Typically, \ac{pcm} requires a relatively high programming current (>$200 \mu A$), but this can be mitigated to less than $10 \mu A$ by scaling down the device size~\cite{pcm-prog1, pcm-prog2}. As \ac{pcm} stores data based on resistance, it can be programmed to encompass more than two resistance states, allowing for multi-level cells to represent more than a single bit of information. Nevertheless, relying on multiple resistance states for prolonged periods poses challenges, as the device resistance tends to drift over time, making it difficult to discern between resistance states.

\subsection{Resistive RAM (\ac{rram})}
\label{subsec:reram}
\ac{rram} is another class of resistive memory technologies that utilizes the resistive switching phenomenon in metal oxide materials to store data~\cite{ReRAM}. As shown in Fig.~\ref{fig:reram}, a typical \ac{rram} cell comprises a top and a bottom electrode with a thin oxide layer sandwiched in between. To achieve resistive switching, a high electric field is applied to the \ac{rram} cell, leading to the creation of oxygen vacancies within the metal oxide layer. This process results in the formation of conductive filaments, causing the device state to transition from a high resistance to a low resistance (set) state. To revert to the high resistance (reset) state, the device is subjected to $V_{RESET}$, which breaks the conductive filament, allowing the oxygen ions to migrate back to the bulk. 
Compared to \ac{pcm}, \ac{rram} exhibits several advantages, including higher write endurance (>$10^{10}$), faster write operations, larger resistance on-off ratios, and improved scalability prospects~\cite{ReRAM}. However, ReRAM does suffer from inconsistent electrical characteristics, meaning it exhibits larger variations in resistance across different devices~\cite{pcm-prog1}.



\subsection{Magnetic RAM (MRAM)} 
\label{subsec:mram}
\Ac{mram} store data in nano-scale ferromagnetic elements via magnetic orientation~\cite{mram-review}. An \ac{mram} cell is a \ac{mtj} device composed of two ferromagnetic layers, namely a fixed reference layer and a free layer, separated by an insulating layer. The free layer holds the data bit, and reading it involves passing an electric current and measuring its resistance. 
For data writing into an \ac{mram} cell, various techniques can be used. The most common method is the \ac{stt}, which utilizes spin-polarized electric current to change the free layer's magnetic orientation. \ac{sot}-\ac{mram}, on the other hand, uses an in-plane current through the heavy metal layer to generate a spin current that exerts a torque on the magnetization of the free layer.
The relative orientations of the free and fixed layers result in different resistance states.
\ac{mram} exhibits virtually unlimited endurance and acceptable access latency. However, it is faced with challenges such as a larger cell size and a smaller on/off resistance ratio, limiting an \ac{mram} cell to store only one bit of data~\cite{FeRAM-1}.

\subsection{Ferroelectric Field-Effect Transistor (FeFET)}
\label{subsec:fefet}
Since the discovery of ferroelectricity in hafnium oxide, \ac{fefet}s have received considerable attention. \ac{fefet}s are non-volatile three-terminal devices, offering high $I_{on}/I_{off}$ ratios and low read voltage. Unlike \ac{mos}-FETs, \ac{fefet}s incorporate a ferroelectric oxide layer in the gate stack, as shown in Figure~\ref{fig:fefet}. The nonvolatility arises from hysteresis due to the coupling between the ferroelectric and CMOS capacitances (C$_{FE}$ and C$_{CMOS}$).
The three-terminal structure of \ac{fefet}s enables separate read and write paths. Reading involves sensing the drain-source current, while writing involves switching the ferroelectric polarization with an appropriate V$_{gs}$ voltage. Unlike two-terminal devices with variable resistance, \ac{fefet}s do not require a drain-source current during the writing process, leading to low writing energy consumption~\cite{cim-fefet}.
There are various \ac{cim} architectures exploiting different properties of \ac{fefet}s. For instance, for boolean operations, \cite{cim-fefet} proposes precharging the bitlines followed by simultaneous activation of the target rows, and using differential sense amplifiers to discern the output. 


\subsection{Comparison and discussion}
\label{subsec:nvms-comp}
Table~\ref{fig:mem-tech-comp} presents a comparison between mainstream and emerging memory devices, discussed in the preceding sections, with respect to performance, reliability, and energy consumption, among others. This analysis gives insights into their suitability for different application domains. It is clear that no single memory device can optimize all metrics. Nonetheless, recent investigations into machine learning use cases show that different phases of machine learning tasks demand different memory device properties, potentially offering the opportunity to employ various devices for various application domains (or tasks within a domain) and achieve the best results~\cite{kendall2020building}.

\begin{table}
\scriptsize
    \begin{center}
    \begin{tabular}{p{0.18\linewidth}p{0.10\linewidth}p{0.10\linewidth}p{0.10\linewidth}
    p{0.10\linewidth}p{0.10\linewidth}p{0.10\linewidth}}
    \toprule
    \textbf{Device} & SRAM      & DRAM & RRAM   & PCM   & STT-MRAM  & FeFET     \\ \midrule
    \textbf{Write time} & $1-10ns$ & $>20ns$  & $>10ns$ & $\sim50ns$ & $>10ns$ & $\sim10ns$ \\ 
    \textbf{Read time} & $1-10ns$ & $>20ns$  & $>10ns$ & $>10ns$  & $>10ns$ & $\sim10ns$ \\ 
    \textbf{Drift} & No & No & Weak & Yes & No & No \\ 
    \textbf{Write energy (per bit)} & $1-10fJ$ & $10-100fJ$ & $0.1-1pJ$ & $100pJ$ & $\sim100 fJ$ & $>1fJ$ \\ 
    \textbf{Density} & Low & Medium & High & High & Medium & High \\ 
    \textbf{Endurance} & $>10^{16}$ &$>10^{16}$ & $>10^{5}-10^{8}$ & $>10^{5}-10^{8}$ & $>10^{15}$  & $>10^{15}$  \\ 
    \textbf{Retention} & Low & Very Low & Medium & long & Medium & long \\
\bottomrule
\end{tabular}
\end{center}
\caption{A comparison of the key features across different mainstream CMOS and emerging memristive technologies~\cite{memristors-cmos}.} 
\label{fig:mem-tech-comp}
\end{table}


\ac{pcm}, \ac{rram}, \ac{mram} and \ac{fefet} fall under the category of memristive technologies, where devices can exhibit multiple resistance states. 
This characteristic has been effectively leveraged to perform \ac{mvm}, as depicted in Figure~\ref{fig:cim-crossbar}. 
Although the analog computation may not be entirely precise, some loss in accuracy is acceptable in many application domains, particularly for machine learning applications. Numerous \ac{cim} architectures have been proposed using this technique to accelerate neural network models (see 
Section~\ref{subsec:cim-arch}).

\begin{figure}[tbh]
\centering
\includegraphics[scale=0.11]{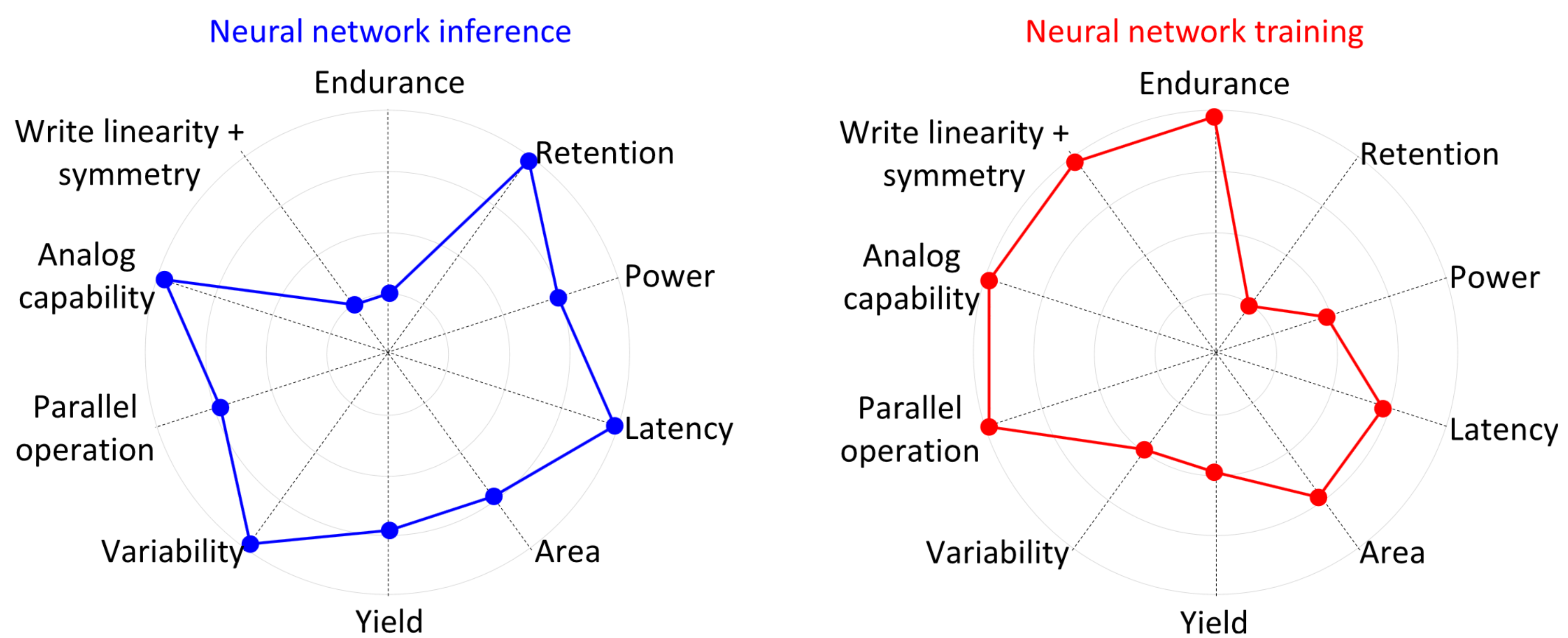}
\caption{A spider chart showing the importance of different attributes of the NVM technologies for the neural network training and inference. More distance from the center means more important~\cite{kendall2020building}.} 
\label{fig:NNtrain-vs-infer}
\end{figure}

Figure~\ref{fig:NNtrain-vs-infer} shows the importance of various device properties for neural network training and inference. 
For training, frequent weight updates within the memory are crucial, making memory technologies like \ac{pcm} and \ac{rram}, with limited endurance and expensive write operations, poorly suitable for training acceleration. However, in inference, where operations are predominantly read-based with minimal writes to the crossbar array, the same technology could outperform others by orders of magnitude. Similarly, retention for training is the least important but is critical for inference. 

\begin{figure}[tbh]
\centering
\includegraphics[scale=0.12]{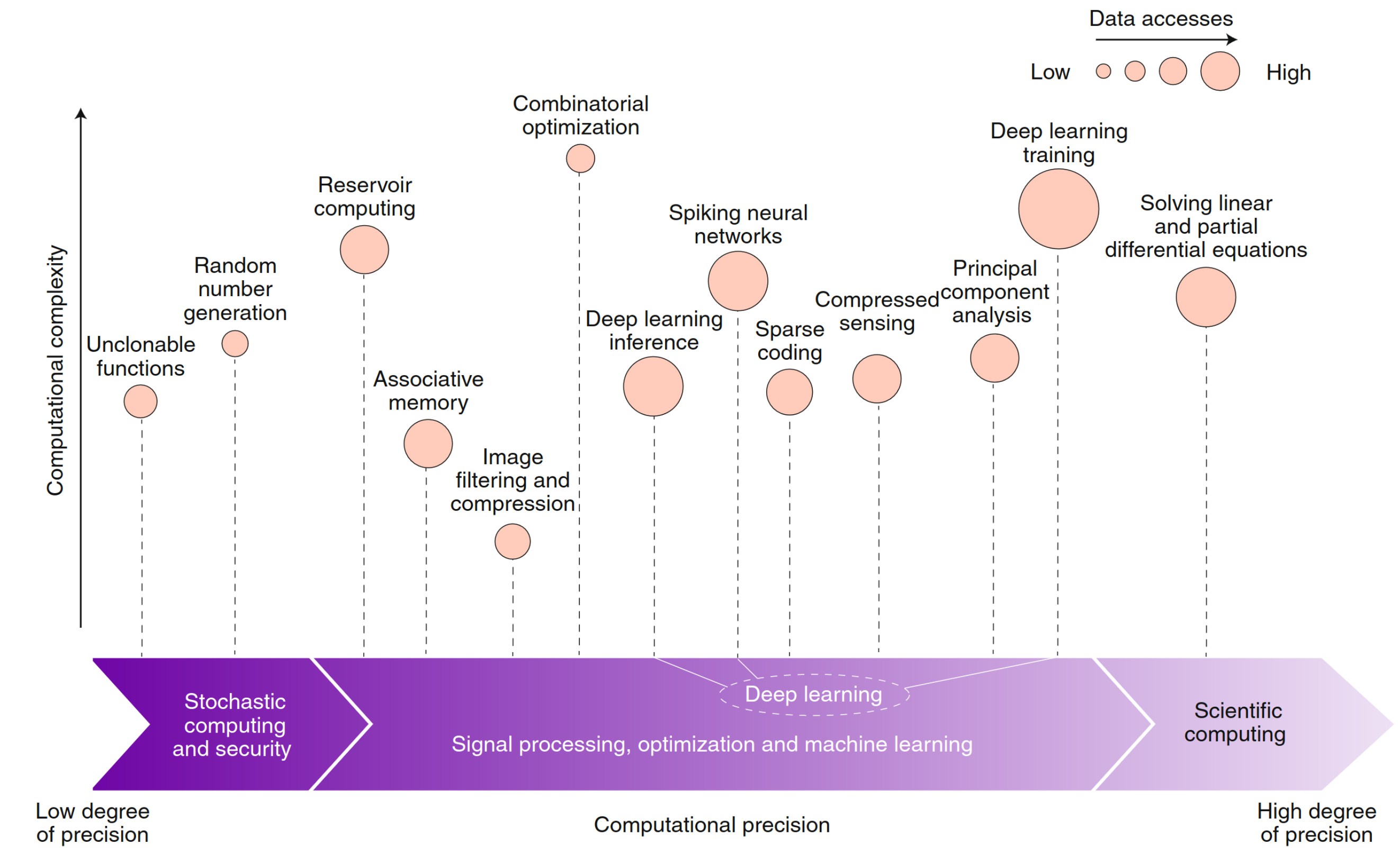}
\caption{The applications landscape for \ac{cim} and \ac{cnm}~\cite{cim_survey}.} 
\label{fig:cim-apps}
\end{figure}

The arguments around Figure~\ref{fig:NNtrain-vs-infer} generally hold true for other metrics and other application domains. Although machine learning constitutes a significant area, it is not the only domain benefiting from the \ac{cim} paradigm. Many other data-intensive application domains have also effectively exploited the \ac{cim} paradigm. Figure~\ref{fig:cim-apps} shows a landscape of \ac{cim} applications, emphasizing precision considerations, computational complexity and memory access requirements. These applications are classified into three categories based on their precision demands. This data pertains to 2020. Over the past two years, additional application domains, such as databases, bioinformatics, and solving more complex algebraic tasks, have gained significant attention as well.

%% file: contents/selected-arch.tex
In this section, we discuss some prevalent \ac{cim} and \ac{cnm} architectures, explaining their programming models and systems integration. It is important to highlight that there exist COM accelerators optimized for specific domains, achieving throughput similar to \ac{cnm}/\ac{cim} counterparts, albeit at the expense of higher energy consumption. These accelerators are beyond the scope of this paper. 

This section is structured as follows: In Section~\ref{subsec:cnm-arch}, we provide an overview of \ac{cnm} systems, starting with academic designs and progressing to commercial \ac{cnm} systems, including both planar 2D and stacked DRAM structures. Section~\ref{subsec:cim-arch} follows a similar organization for \ac{cim} systems employing various technologies. In Section~\ref{subsec:compare}, we conduct a comparative analysis of different \ac{cim}/\ac{cnm} systems, while Section~\ref{subsec:challenges} outlines the key challenges faced by these innovative architectures.

\subsection{CNM architectures}
\label{subsec:cnm-arch}
The core principle of compute-near-memory is to perform computations in the memory proximity by placing \acp{pu} on/near the memory chip. The first \ac{cnm} architecture dates back to the 1990s that aimed at integrating compute units with embedded DRAM on the same chip to achieve higher bandwidth. However, due to technological limitations and costly fabrication processes, even the promising initial \ac{cnm} proposals like IRAM~\cite{IRAM}, DIVA~\cite{DIVA}, and FlexRAM~\cite{flexRAM} never commercialized.

In recent years, due to the advancements in integration and die-stacking technologies, \ac{cnm} has regained interest within both industry and academia. \acp{pu} are being integrated at different locations within memory devices, including within the memory chip as well as outside the memory chip on the module level, i.e., dual in-line memory module (DIMM). A DIMM typically consists of multiple memory chips, each consisting of multiple ranks, banks, and subarrays. It is worth noting that some researchers also classify \ac{pu}s integrated at the memory controller level as \ac{cnm}. However, following the classification and terminology adopted in this report, we categorize it under the COM class.

In the following, we explain some of the common \ac{cnm} architectures. We start by examining the planar 2D DRAM-based \ac{cnm} designs, then transition to discussing the 2.5D and 3D DRAM-based \ac{cnm} systems, and ultimately conclude on the NVM-based \ac{cnm} architectures.

\subsubsection{The UPMEM system}
\label{sss:upmem}
\begin{figure}[tbh]
\centering
\includegraphics[scale=0.12]{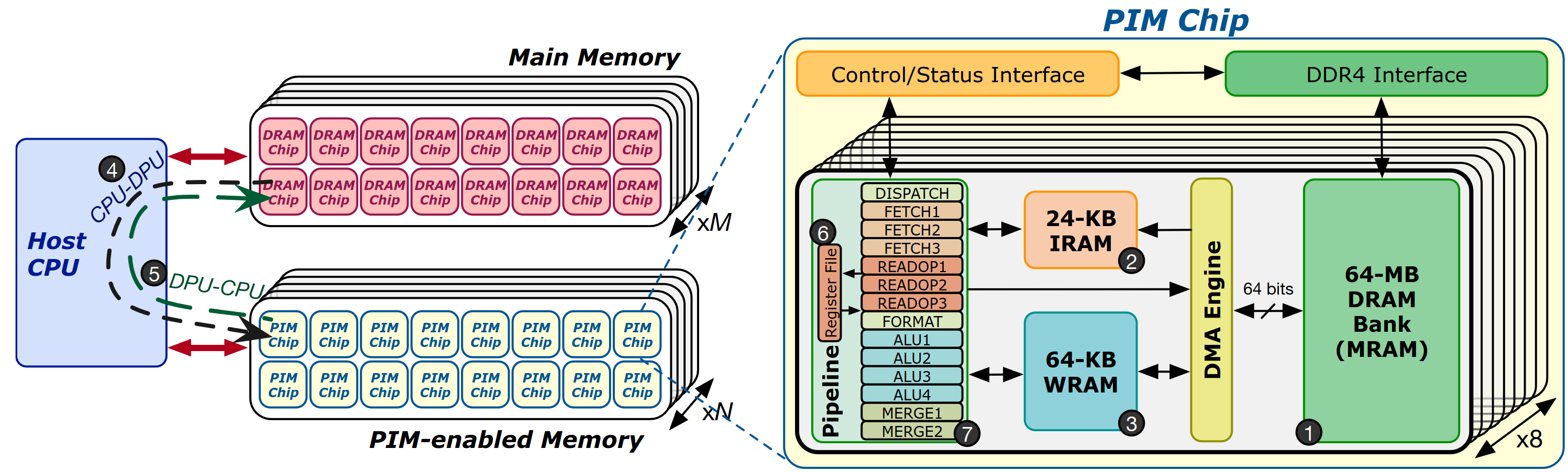}
\caption{An overview of the UPMEM architecture~\cite{upmembenchmarking}.} 
\label{fig:upmem}
\end{figure}

UPMEM is a recent commercial near-bank \ac{cnm} system and is publicly available~\cite{upmem}. Figure~\ref{fig:upmem} gives a detailed overview of the UPMEM architecture. The memory modules are divided into PIM-enabled memory and main memory (conventional). The PIM-enabled memory combines co-processors known as \acp{dpu} with conventional DDR4 DRAM on the same die. 

\acp{dpu} are 32-bit general-purpose RISC processors, comprising a 64kB SRAM-based scratchpad working memory known as WRAM, a 24kB SRAM-based instruction memory referred to as IRAM, and a shared main memory named MRAM, based on DRAM technology. As shown in the figure (lower left), each DIMM consists of 16 memory chips, with each chip housing 8 banks, and each bank containing one \ac{dpu}. The latest UPMEM systems can support up to 20 DIMMs. 
~\\\noindent{\textbf{\ac{dpu}-\ac{dpu} communication:}} 
\ac{dpu}s in UPMEM can have up to 24 hardware threads called tasklets. Within the same \ac{dpu}, tasklets can share data through MRAM and WRAM, however, \ac{dpu}s can not communicate with each other directly and must go through the host for any possible data sharing. 
~\\\noindent{\textbf{Progammability:}}
For programmability, UPMEM offers its own \ac{sdk} consisting of an UPMEM compiler and runtime libraries. \ac{dpu} programs are written in the C language including specific library calls. 
The runtime library provides functions for data and instruction transfers between different memory, e.g., MRAM-IRAM, MRAM-WRAM etc.; executing various functions on the \ac{dpu}s; and synchronization (mutex locks, barriers, handshakes, and semaphores). 

Although UPMEM claims they have an easily programmable \ac{sdk}, programming the system has several challenges. The programmer is responsible for efficient mapping of executions and load-balancing on thousands of \ac{dpu}s, managing data transfer, and ensuring coherence of data between CPU and \ac{dpu}s. 

\subsubsection{CNM for \ac{mvm} in DRAM}
\label{sss:mvid-mcdram}
McDRAM~\cite{McDRAM} and MViD~\cite{mvid} (both involving Samsung Electronics) aimed at accelerating machine learning workloads by embedding \ac{mac} units within the DRAM bank. Similar to UPMEM, both McDRAM and MViD utilize 2D DRAM (LPDDR in these cases) and incorporate \acp{pu} within the memory banks to exploit the higher internal memory bandwidth. However, unlike UPMEM, these architectures are domain-specific and hence employ fixed functional units (\ac{mac}s) instead of general-purpose programmable cores.

\begin{figure}[tbh]
\centering
\includegraphics[scale=0.10]{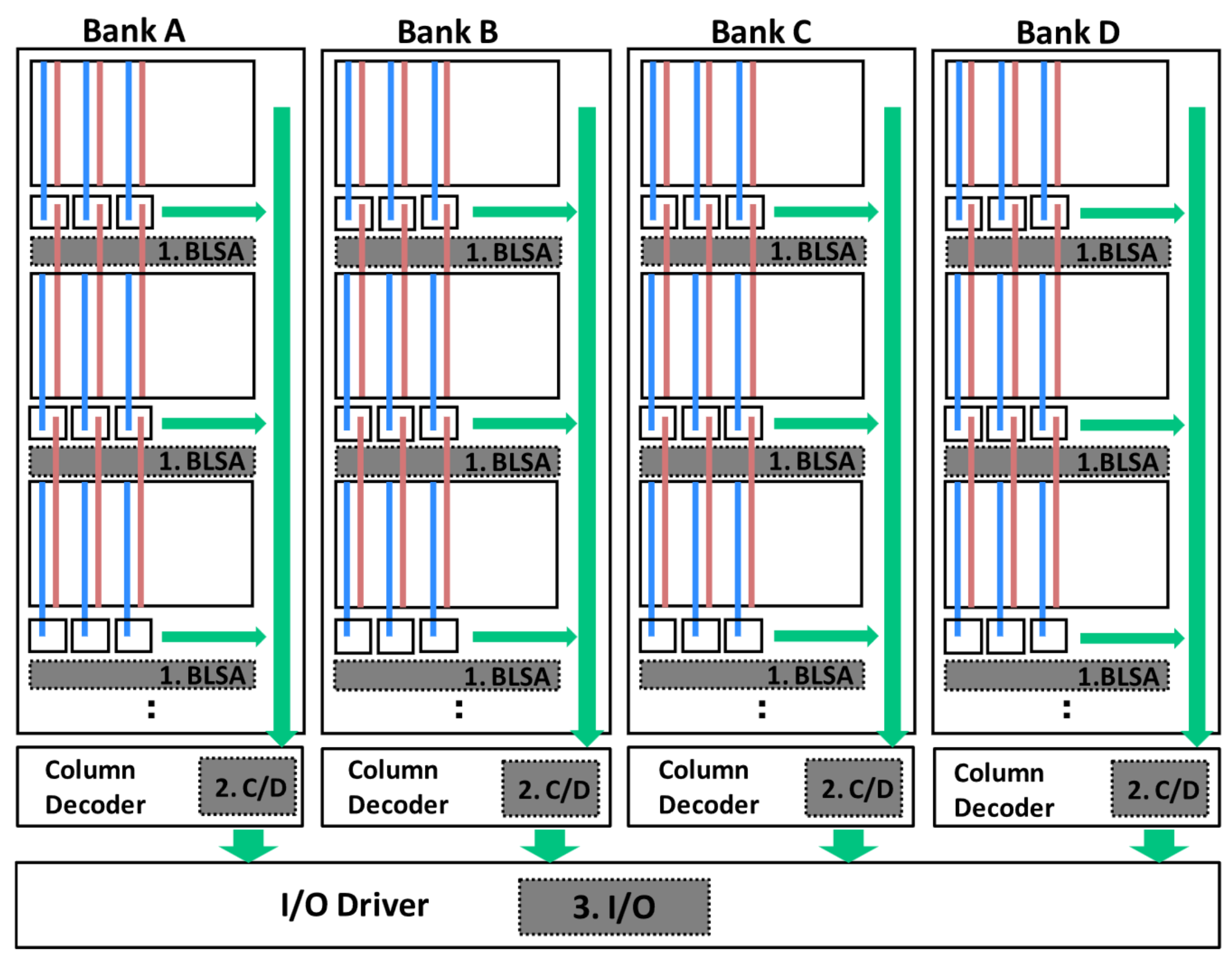}
\caption{The McDRAM architecture with three possible locations for MAC units~\cite{McDRAM}.} 
\label{fig:mcdram}
\end{figure}

Figure~\ref{fig:mcdram} shows the McDRAM architecture along with the three locations (column decoder, bitline \ac{sa}s, and I/O drivers) where MAC units were employed and evaluated. Each McDRAM chip consists of 4 banks and each bank has four 8-bit MAC units. 
The multiplication is performed in parallel by multiplying rows of the matrix with the input vector. 

~\\\noindent\textbf{Programmability: }
McDRAM is a fixed-function accelerator and offers a single interface function (matmul) that triggers the device driver to configure the control registers of the memory controller.
It operates in two modes, memory and compute modes, determined by a configuration register. In compute mode, McDRAM performs \ac{mvm} tasks. For the management of \ac{mvm} within compute mode, it introduces six novel DRAM commands that leverage existing DRAM I/O signals, rendering no modifications to DRAM I/O signals.
McDRAM, a fixed-function accelerator, employs a single interface function (matmul) triggering the device driver to configure memory controller control registers. It operates in two modes, memory and compute, determined by a configuration register. In compute mode, McDRAM performs \ac{mvm} introducing six novel DRAM commands for \ac{mvm} management without modifying existing DRAM I/O signals.

\begin{figure}[tbh]
\centering
\includegraphics[scale=0.11]{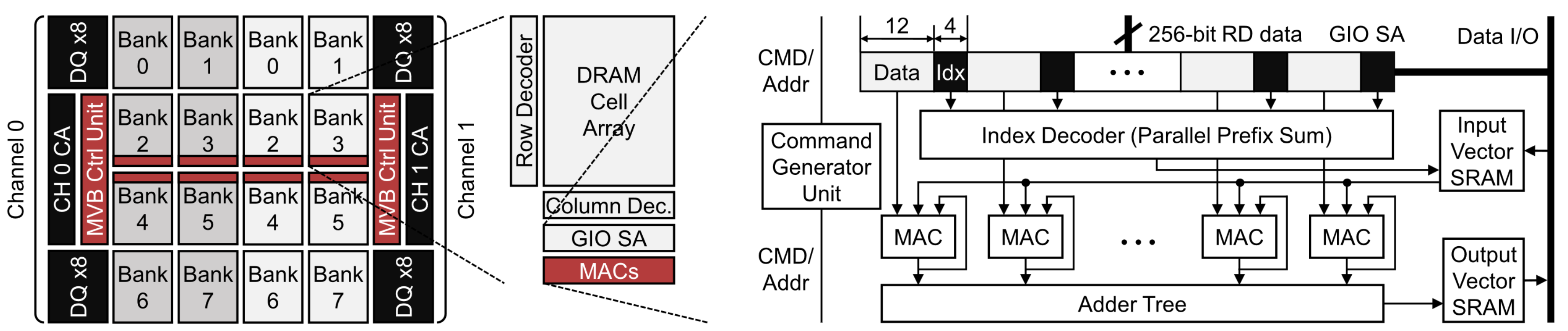}
\caption{The MViD architecture~\cite{mvid} where each bank has 16 MAC units.} 
\label{fig:mvid}
\end{figure}

The MViD architecture depicted in Figure~\ref{fig:mvid} is similar to the McDRAM design and is specifically optimized for edge devices. Much like McDRAM, MViD incorporates MAC units within the DRAM I/O drivers to capitalize on the internal bandwidth of the DRAM. However, unlike McDRAM, MViD introduces a partitioning of memory banks into two categories: \ac{mvm} banks that are equipped for MAC units and two SRAM structures to hold the input and output vectors and non-\ac{mvm} banks (traditional). This division enables concurrent access to both types, meaning that multiplication operations in \ac{mvm} banks can occur simultaneously with the CPU accesses to the non-\ac{mvm} banks.

\subsubsection{Samsung's CNM systems}
\label{sss:fimdram}
Samsung is probably ahead of everyone in the race for commercial \ac{cnm} systems. In the following, we discuss two of their recent promising (and complete) solutions. 

~\\\noindent{\textbf{PIM-HBM:}}
Samsung has recently introduced a \ac{cnm} architecture referred to as Function-in-Memory DRAM (FIMDRAM)\cite{samsung-ISSCC} or PIM-HBM\cite{samsung-ISCA}. It incorporates 16 \ac{simd} engines within the memory banks, enabling bank-level parallelism. As reported in~\cite{samsung-ISCA}, their design does not disrupt crucial elements on the memory side, such as the sub-array and bank in conventional DRAM, making its integration seamless and straightforward. Importantly, it does not require any modifications to contemporary commercial processor components, including DRAM controllers. It is designed for host processors to manage PIM operations via standard DRAM interfaces. This feature allows for a straightforward substitution of existing JEDEC-compliant DRAM with PIM-DRAM across various systems.

\begin{figure}[tbh]
\centering
\includegraphics[scale=0.11]{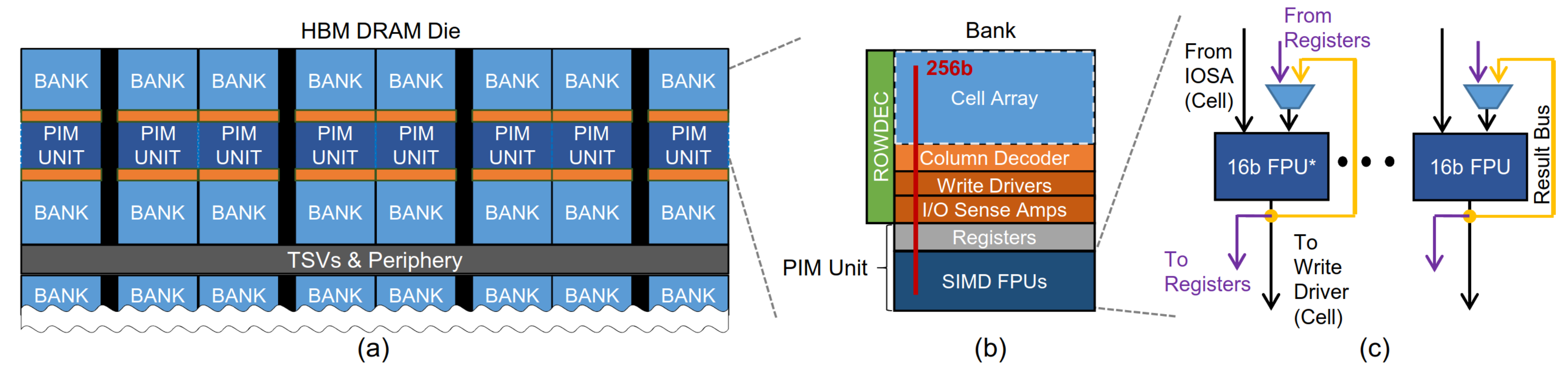}
\caption{Samsung's PIM-HBM (a) HBM die organization (b) Bank coupled with a PIM unit (c) PIM unit data path~\cite{samsung-ISCA}.} 
\label{fig:fimdram}
\end{figure}

While Samsung reports that their design is compatible with any DRAM family, they have showcased its functionality using the 2.5D high bandwidth memory (HBM) DRAM. Figure~\ref{fig:fimdram} provides a high-level view of this architecture. Each bank comprises 16 \ac{simd} \acp{fpu}, with each FPU consisting of a 16-bit floating-point adder and a 16-bit floating-point multiplier. Furthermore, each FPU is equipped with data registers (GRFs), control and instruction registers (CRF, SRF), and an internal control unit. The internal control unit orchestrates operation sequences without necessitating modifications to the memory controller. When operating in PIM mode, the PIM execution units within all banks simultaneously respond to a standard DRAM column (Read or Write) command initiated by the host processor and execute a wide SIMD operation with deterministic latency in a lock-step manner. 

~\\\noindent{\textbf{Programmability: }} PIM-HBM comes with an \ac{isa}, a software stack, and a specific programming model. 
The software stack presents a native execution path that does not require any modifications to the input code. The framework takes the high-level representation of an application and transforms it into device code. Furthermore, it offers a direct execution path that permits direct invocation of various function calls using the ``PIM custom op``. The PIM runtime includes a collection of modules responsible for tasks like operations offloading, memory allocation, and execution on the FPUs.

HBM-PIM is a commercial accelerator and, as per, Samsung is already used by companies. Here is an excerpt from Samsung's newsroom:

\textit{“Xilinx has been collaborating with Samsung Electronics to enable high-performance solutions for data center, networking, and real-time signal processing applications starting with the Virtex UltraScale+ HBM family, and recently introduced our new and exciting Versal HBM series products,” said Arun Varadarajan Rajagopal, senior director, Product Planning at Xilinx, Inc. “We are delighted to continue this collaboration with Samsung as we help to evaluate HBM-PIM systems for their potential to achieve major performance and energy-efficiency gains in AI applications.”}

\begin{figure*}[hbt!]
\subfloat[Samsung-Facebook AxDIMM hardware module and architecture~\cite{samsung-AxDIMM}\label{fig:axDIMM}]{%
        \begin{minipage}{0.38\columnwidth}
\includegraphics[scale=0.078]{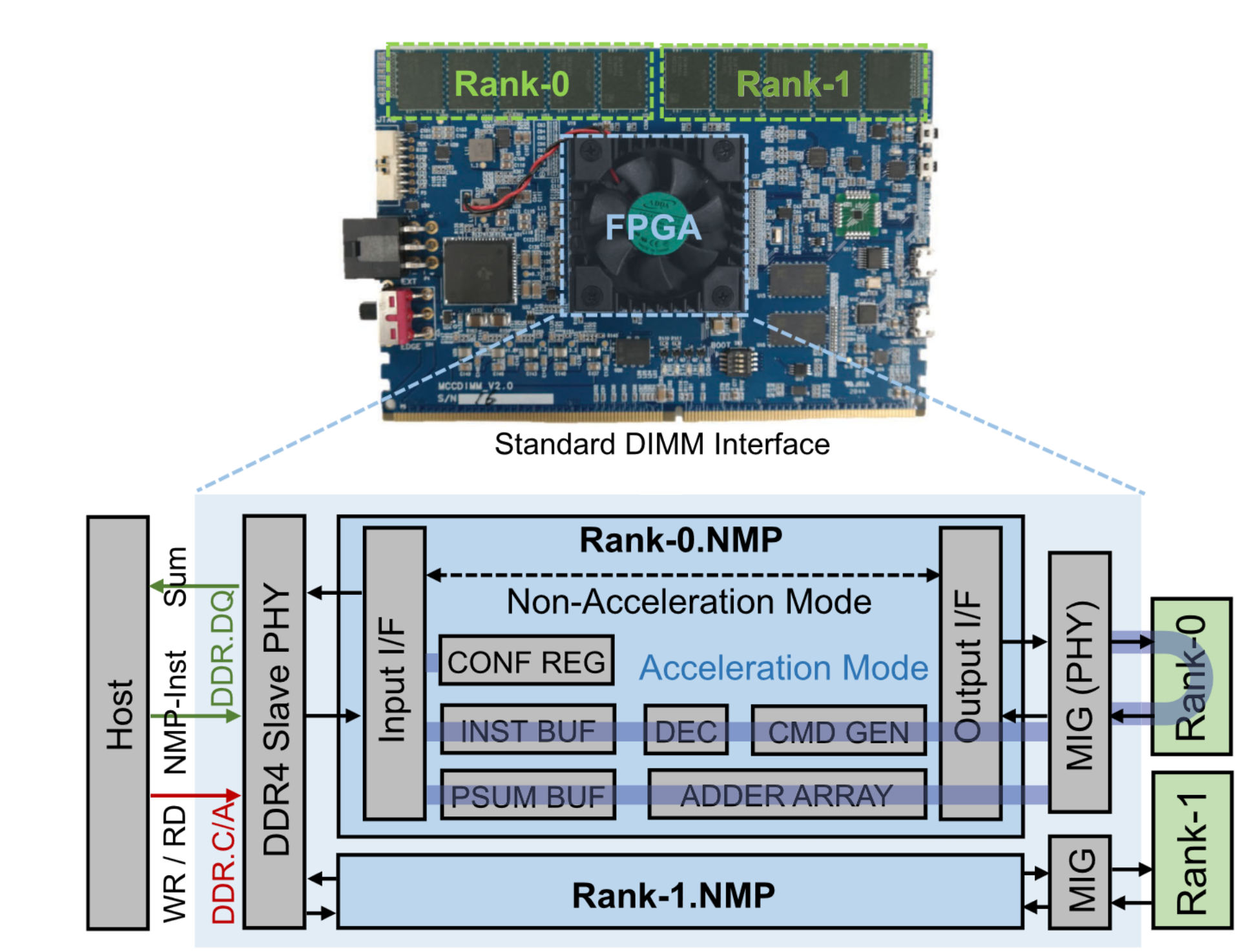}
        \end{minipage}%
    }
    \hfill
     \subfloat[AiM architecture~\cite{aim-ISSCC}. \label{fig:aim}]{%
        \begin{minipage}{0.6\columnwidth}
\includegraphics[scale=0.06]{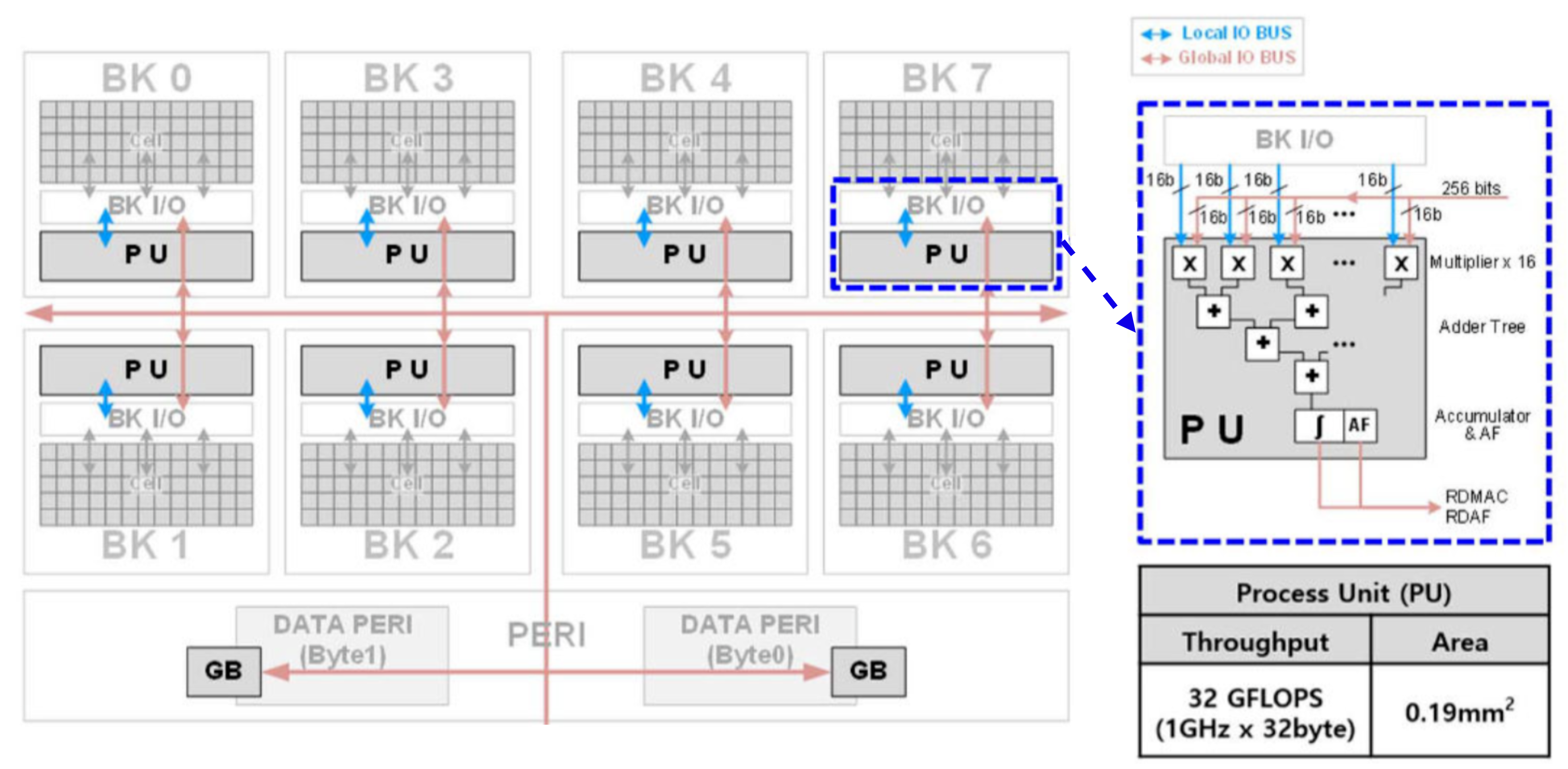}
        \end{minipage}%
    }
     \caption{PUMA tile and core architectures~\cite{puma}.}
     \label{fig:puma}
\end{figure*}

~\\\noindent{\textbf{AxDIMM (by Samsung-Facebook:)}}
Samsung is also working on the development of an FPGA-enabled \ac{cnm} platform named AxDIMM. In collaboration with Facebook, this solution has showcased its effectiveness in a personalized recommender system. As shown in Figure~\ref{fig:axDIMM}, the \ac{cnm} architecture (RANK) is the same as Samsung's HBM-PIM, but the controlling unit is FPGA that starts the execution, maps computations to the RANK, and gets back the results. Like the HBM-PIM, AxDIMM has a complete software stack that allows programming the architecture without changing the input code or manually writing code using AxDIMM python API.

For this product, Samsung also seems to be in discussion with SAP HANA. Here is another excerpt from the newsroom: 

\textit{``SAP has been continuously collaborating with Samsung on their new and emerging memory technologies to deliver optimal performance on SAP HANA and help database acceleration,” said Oliver Rebholz, head of HANA core research \& innovation at SAP. “Based on performance projections and potential integration scenarios, we expect significant performance improvements for in-memory database management system (IMDBMS) and higher energy efficiency via disaggregated computing on AXDIMM. SAP is looking to continue its collaboration with Samsung in this area”}.

\subsubsection{SK hynix's accelerator-in-memory}
\label{sss:aim}
SK hynix's accelerator-in-memory (AiM) is another \ac{cnm} system that targets the machine learning application domain~\cite{aim, aim-ISSCC}. As stated in~\cite{aim-ISSCC}, ``Samsung's FIMDRAM is near commercialization, but the required HBM technology may prevent it from being applied to other applications due to its high cost''. AiM fundamentally follows a very similar design approach to Samsung's FIMDRAM but utilizes GDDR6 instead. 

Figure~\ref{fig:aim} provides an overview of the AiM architecture. As depicted, each bank is equipped with a processing unit (PU) that executes a MAC operation using 16 multiplier units and an adder tree. The adder tree can be deactivated for operations not requiring additions. Similar to the FIMDRAM design, pairs of banks can establish direct communication. For inter-group communication, an internal 2KB SRAM structure within the periphery facilitates the process.

Although the programming model is not explicitly explained, the presented set of commands in AiM implies an interface enabling interaction with the device for various operations. Some of these operations are particularly interesting, such as the ability to perform computations within banks of different granularities (1, 4, 16) and data movement functions that can be utilized to implement row-cloning within DRAM.

\subsubsection{AxRAM}
\label{sss:axRAM}
AxRAM targets optimizing for the off-chip memory communication bottleneck in GPUs by integrating approximate MAC units in the DRAM~\cite{axRAM}. The fundamental idea is to exploit the inherent approximability of numerous GPU applications and perform approximate calculations directly within the memory banks, thereby reducing data movement and energy consumption. AxRAM leverages the concept of neural transformation, a technique that accelerates a wide range of applications by approximating specific sections of GPU code and transforming them into a neural representation composed primarily of \ac{mac} and \ac{lut} operations for nonlinear function calculation. The multiplications in the MAC operations are further approximated with limited iterations of shift-add and LUT accesses. These approximate units are connected to the wide data lines that connect the DRAM banks to the global I/O, keeping the banks and memory column unchanged.

\begin{figure}[tbh]
\centering
\includegraphics[scale=0.18]{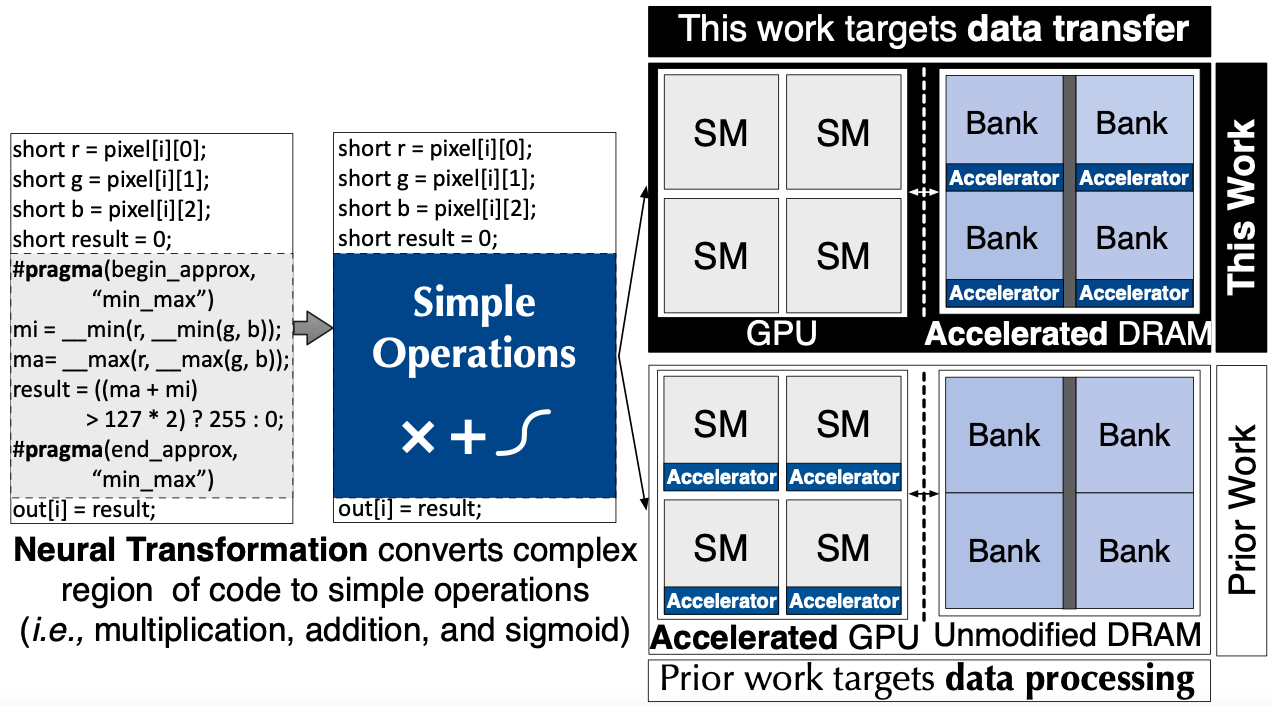}
\caption{Transformation of complex code into simple operations (left) and the AxRAM architecture compared to state-of-the-art (right)~\cite{axRAM}.} 
\label{fig:axRAM}
\end{figure}

Figure~\ref{fig:axRAM} shows a sample example where the GPU code is transformed into MAC and lookup operations. Once such patterns are identified and transformed, they are offloaded to the in-DRAM accelerator. The new instructions that invoke and configure the in-DRAM accelerators are added to the GPU's ISA and are exposed to the compiler. As for the flow of execution, initially, all data is assumed to be in one memory chip. The GPU starts normal execution, and for the identified approximate regions, the GPU warps send an initiation request to the on-chip memory controller. The additional logic in the memory controller first sends an invalid signal to the on-chip caches (to ensure data consistency) and subsequently drives the in-DRAM accelerator to perform the computations and store the results in the designated location. To check whether the execution is completed, the memory controller periodically checks the memory-mapped mode register of the DRAM, which is updated by the accelerator. Once the controller detects that this register is set, it signals the GPU that execution is finalized, allowing the GPU to proceed with precise execution of the subsequent instructions.  

\subsubsection{CNM systems based on 3D-stacked DRAM}
\label{sss:3d-stacked}
All the \ac{cnm} architectures discussed so far (except FIMDRAM) are based on planar 2D DRAM. However, the resurgence in \ac{cnm} systems is also primarily attributed to HBM and HMC technologies that seamlessly combine logic and memory within the same package. There is a series of proposals for \ac{cnm} systems leveraging these technologies. In the following, we discuss some of the prominent architectures. 

~\\\noindent\textbf{
TESSERACT}~\cite{3d-cnm-onur} targets accelerating graph-based applications. Their design comprises a host processor and an HMC with multiple vaults, each housing an out-of-order processor. These processors exclusively access their local data partition, while inter-communication is achieved through a message-passing protocol. The host processor, however, can access the complete address space of the HMC. To capitalize on the substantial memory bandwidth available, they introduce prefetching mechanisms.
~\\\noindent\textbf{
TOP-PIM}~\cite{3d-cnm-top-pim} is an architecture that proposes an \ac{apu}. Each APU integrates a GPU and a CPU on the same silicon die. These APUs are linked through high-speed serial connections to several 3D-stacked memory modules. APUs are general-purpose and support a series of applications ranging from graph processing to fluid and structure dynamics. The architecture allows code portability and easy programmability. 
~\\\noindent\textbf{
Active memory cube (AMC)}~\cite{3d-cnm-amc} is also built upon HMC and proposes ``lanes'' in the HMC vault. Each AMC lane consists of a register file, a computational unit, and a load/store unit to support memory accesses. Communication among AMCs is only possible via the host processor. AMC also offers a compiler based on OpenMP for C/C++ and FORTRAN. 
~\\\noindent\textbf{
Heterogeneous reconfigurable logic (HRL)}~\cite{3d-cnm-hrl} leverages the logic layer in the 3D stacked HMC to implement heterogeneous coarse-grained (CGRAs) and fine-grained (FPGAs) logic blocks. 
The architecture separates routing networks for control and data signals, employing specialized units to efficiently handle branch operations and non-uniform data layouts commonly found in analytics workloads.

\subsection{\ac{cim} architectures}
\label{subsec:cim-arch}
\begin{figure}[tbh]
\centering
\includegraphics[scale=0.10]{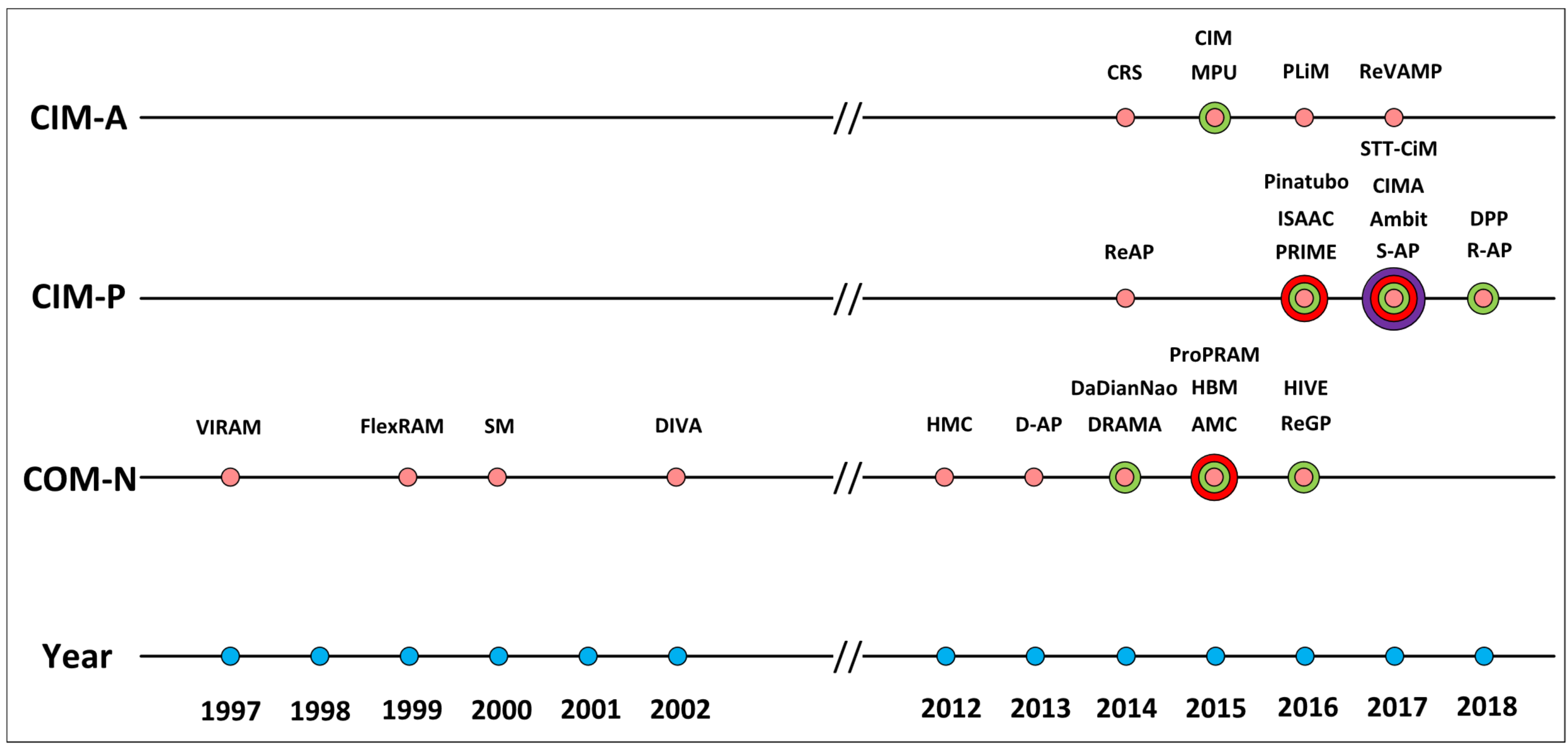}
\caption{A partial timeline of the evolution of \ac{cim} systems (data until 2018)~\cite{delft-cim-survey}. The radius of the circle is proportional to the amount of papers published that year.} 
\label{fig:cim-evolve}
\end{figure}

Much like \ac{cnm}, the concept of \ac{cim} systems is not entirely novel; however, it has gained significant momentum due to breakthroughs in various NVM devices over the past decade. Figure~\ref{fig:cim-evolve} shows a partial landscape of \ac{cim} systems, along with a corresponding timeline. Most of the depicted \ac{cim} accelerators originate from academia and are not taped out. However, in recent years, several semiconductor industry giants, including Intel, Samsung, TSMC, GlobalFoundries, and IBM, have invested in developing their own \ac{cim} prototypes, mostly focused on the machine learning case. IBM, in particular, stands out among others when it comes to the development of \ac{cim} systems for different use cases. 

In this section, we overview some of the prominent \ac{cim} designs from academia and industry. However, before going into the details of individual \ac{cim} designs, we first introduce circuits that are typically used as basic \ac{cim} primitives in these architectures.

\subsubsection{CIM primitives}
\label{sss:cim-primitives}
Each of the \ac{cim} architectures discussed in the following sections is either based on a crossbar, content-addressable-memory, or a boolean and arithmetic logic unit. In the following, we explain all three of them. 

\begin{figure*}[hbt!]
    \subfloat[\ac{cim} crossbar. \label{fig:cim-crossbar}]{
        \begin{minipage}{0.25\columnwidth}
             \centering
            \includegraphics[scale=0.45]{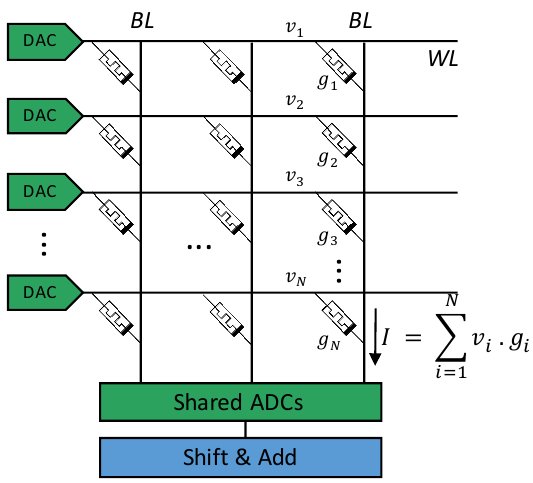}
            
        \end{minipage}%
    }
    \hfill
    \subfloat[\ac{cim} boolean logic. \label{fig:cim-bool}]{
        \hspace{0.5cm}
        \begin{minipage}{0.25\columnwidth}

            \includegraphics[scale=0.45]{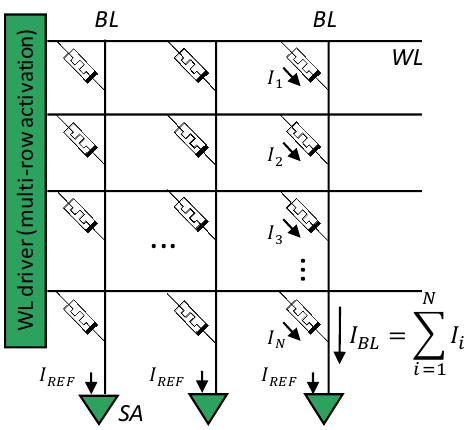}
            \centering
            \vspace{0.03cm}
        \end{minipage}%
    }
    \hfill
    \subfloat[\ac{cim} CAM structure. \label{fig:cim-cam}]{
        \begin{minipage}{0.4\columnwidth}
            \includegraphics[scale=0.35]{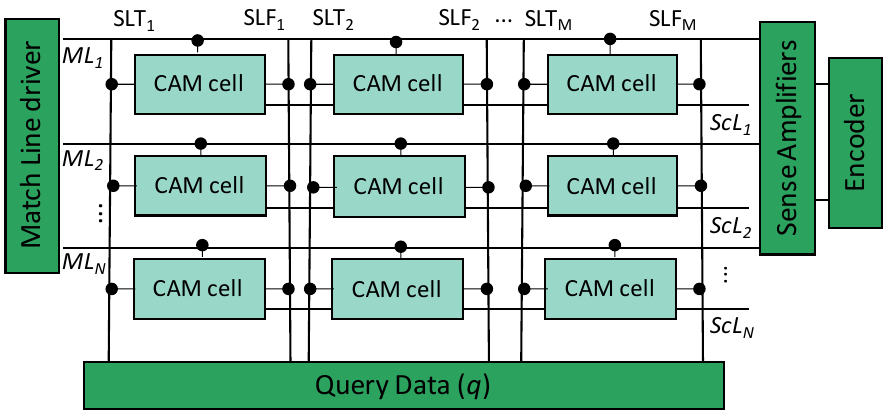}
            \centering
            \vspace{0.5cm}
        \end{minipage}%
    }

     \caption{Fundamental \ac{cim} primitives~\cite{reis2023memory}.}
     \label{fig:cim-primitives}
\end{figure*}

~\\\noindent\textbf{Crossbar:} 
A crossbar is a \ac{cim} configuration in which each input connects to every output through cross-points, comprising memory cells and selectors. Figure~\ref{fig:cim-crossbar} shows a technology-independent crossbar configuration. As we will see in the following sections, crossbars are particularly useful for the machine learning domain as they can compute \ac{mvm} in constant time.
~\\\noindent\textbf{CAM:} 
\ac{cam} is associative memory that enables parallel searches for a given query (input) across all stored content within a CAM array. CAMs are used in pattern matching and search operations from various application domains including databases, networking, and machine learning~\cite{reis2023memory}. Figure~\ref{fig:cim-cam} shows a technology-independent $3\times3$ CAM structure.
~\\\noindent\textbf{Boolean and arithmetic logic in CIM:} 
In this class of \ac{cim}, the \ac{cim} array facilitates a specific set of general operations, such as Boolean logic and arithmetic, to be executed using customized peripheral circuits integrated within the random-access memory (RAM). The operands need to be stored in different rows of an array in a column-aligned fashion where each column represents a bit position. For a \ac{cim} operation, multiple rows are typically activated simultaneously, and the output is sensed and inferred by the peripherical circuitry~\cite{li2016pinatubo,xie2017scouting}. Figure~\ref{fig:cim-bool} shows a technology-independent structure implementing boolean logic. 

\subsubsection{ISAAC (by Hewlett Packard Enterprise)}
\label{sss:isaac}
\begin{figure}[tbh]
\centering
\includegraphics[scale=0.08]{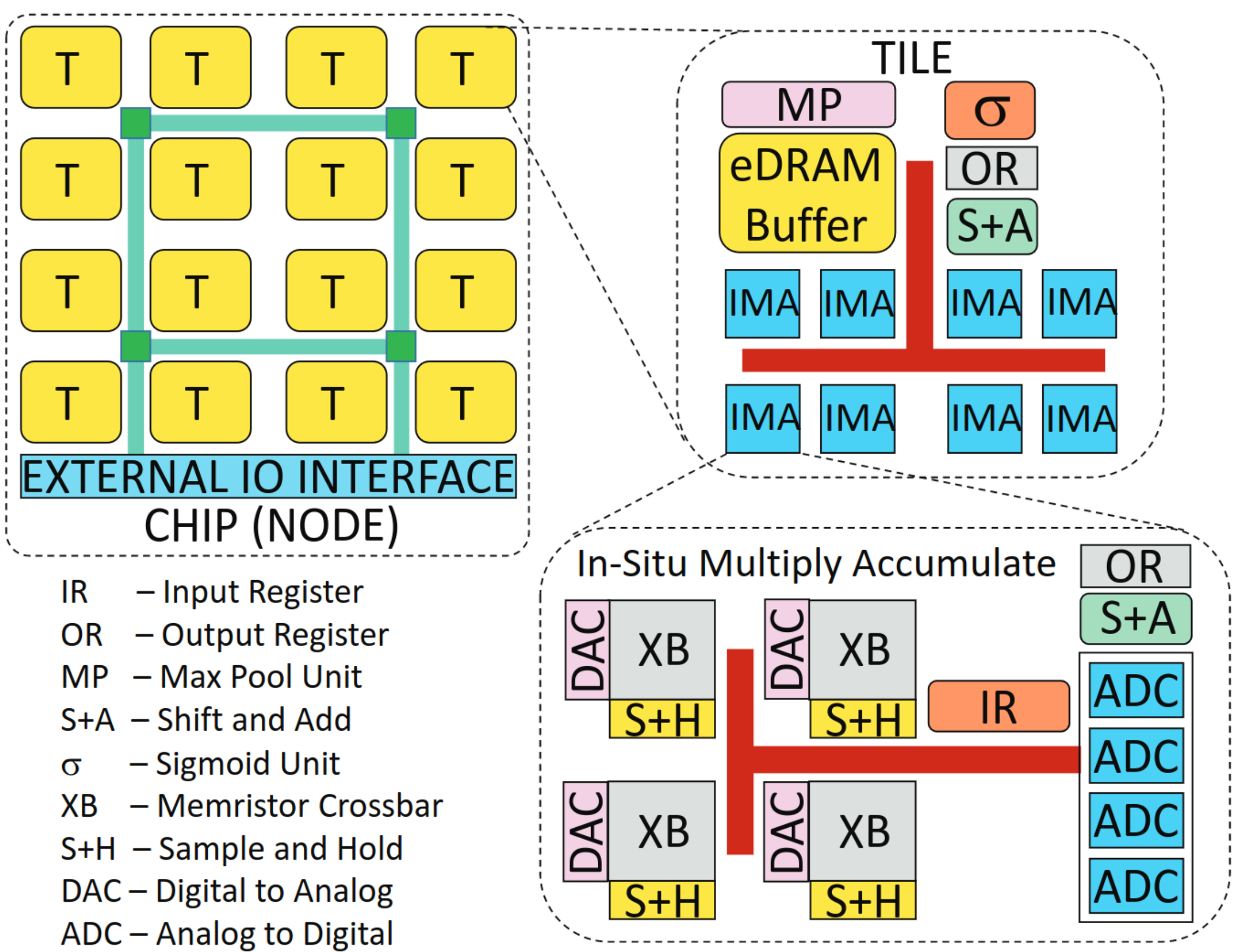}
\caption{ISAAC architecture hierarchy~\cite{isaac}.} 
\label{fig:isaac}
\end{figure}

In-situ analog arithmetic in crossbars (ISAAC)~\cite{isaac} is among the first \ac{cim} accelerators with a complete design targeting \ac{cnn} in RRAM. As shown in Figure~\ref{fig:isaac}, ISAAC's architecture consists of multiple interconnected tiles via a concentrated-mesh (c-mesh) network. Each tile consists of 12 in-situ multiply-and-accumulate (IMA) units, a shift-and-add (S\&A) unit, two sigmoid units, one max-pooling unit, an embedded DRAM (eDRAM) buffer for input data storage and an output register (OR) to accumulate (partial) results. Each IMA integrates its own input register (IR), output register, S\&A units, and eight 128 × 128 resistive crossbar arrays, also abbreviated XB or XBars, that share analog-to-digital converters (ADCs). Each XBar performs analog \ac{mvm} (see Figure~\ref{fig:cim-crossbar}) and is also equipped with a digital-to-analog converter (DAC) and an S\&H circuitry. Communication within a tile is facilitated by a 32-bit inter-tile link.

The ISAAC design uses dataflow pipelining to optimize IMA unit utilization and reduce buffering requirements. Depending on the network's size, each \ac{cnn} layer is mapped to one or multiple IMAs or tiles. Initially, input data is acquired through an I/O connection and stored within a tile's eDRAM buffer. Before being fed to ReRAM XBars within each IMA, the data goes through DACs. Once processed by XBars, the generated feature maps are converted back to digital form and forwarded to max-pooling and activation units. The outcome of the NN layer is then accumulated within the S\&A and OR units and subsequently written to a new eDRAM buffer (for the following layer). The depth of the pipeline corresponds to the depth of the neural network, which presents challenges when training \acp{dnn}. Thus, ISAAC is specifically designed for inference and is not used for training. ISAAC has no mention of the design tools and programming interface.

\subsubsection{PUMA (by Hewlett Packard Enterprise)}
\label{sss:puma}
\begin{figure*}[hbt!]
\subfloat[PUMA's tile architecture\label{fig:puma-tile}]{%
        \begin{minipage}{0.35\columnwidth}
         \includegraphics[scale=0.07]{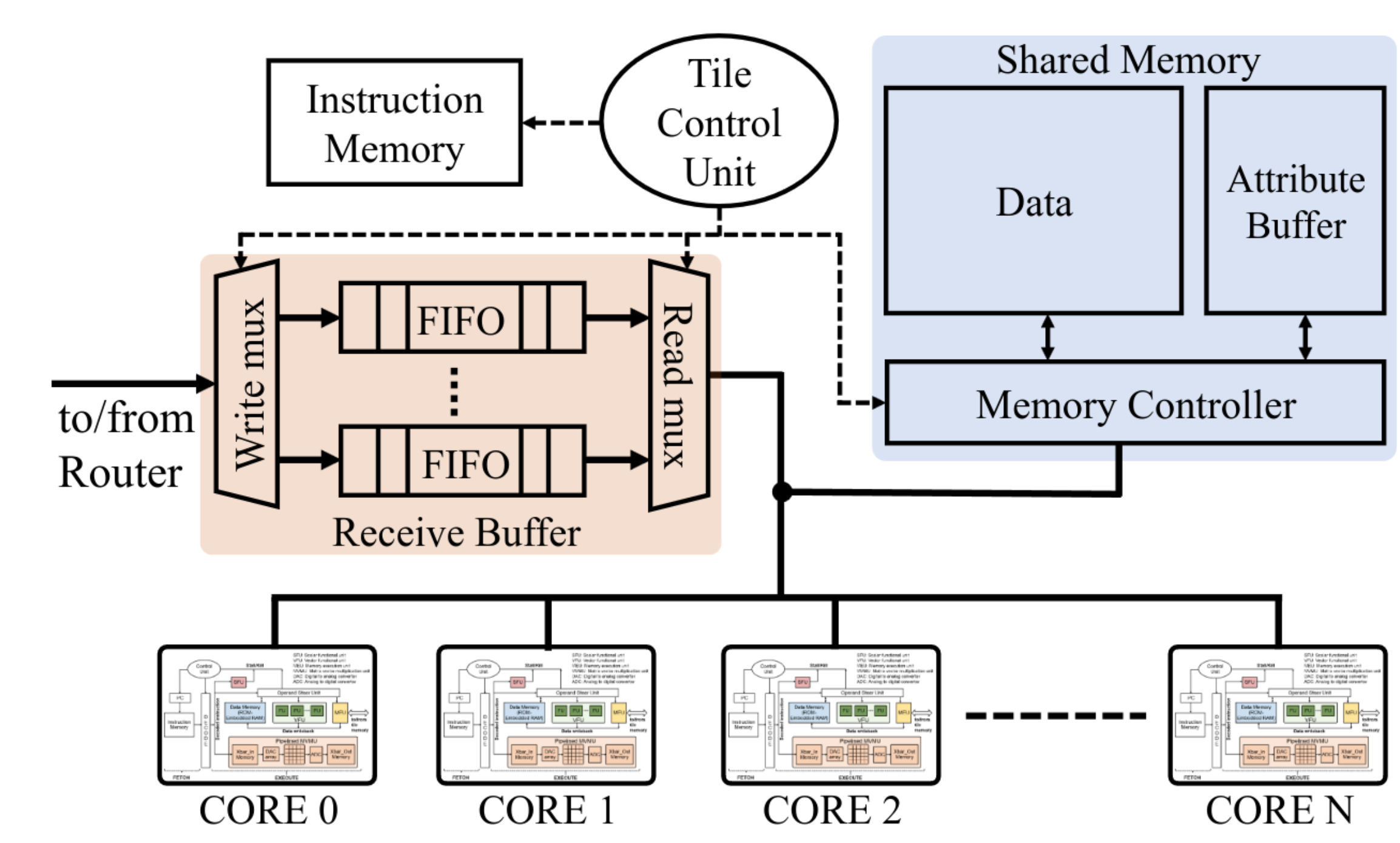}
        \end{minipage}%
    }
    \hfill
     \subfloat[PUMA's core architecture \label{fig:puma-core}]{%
        \begin{minipage}{0.57\columnwidth}
                \includegraphics[scale=.075]{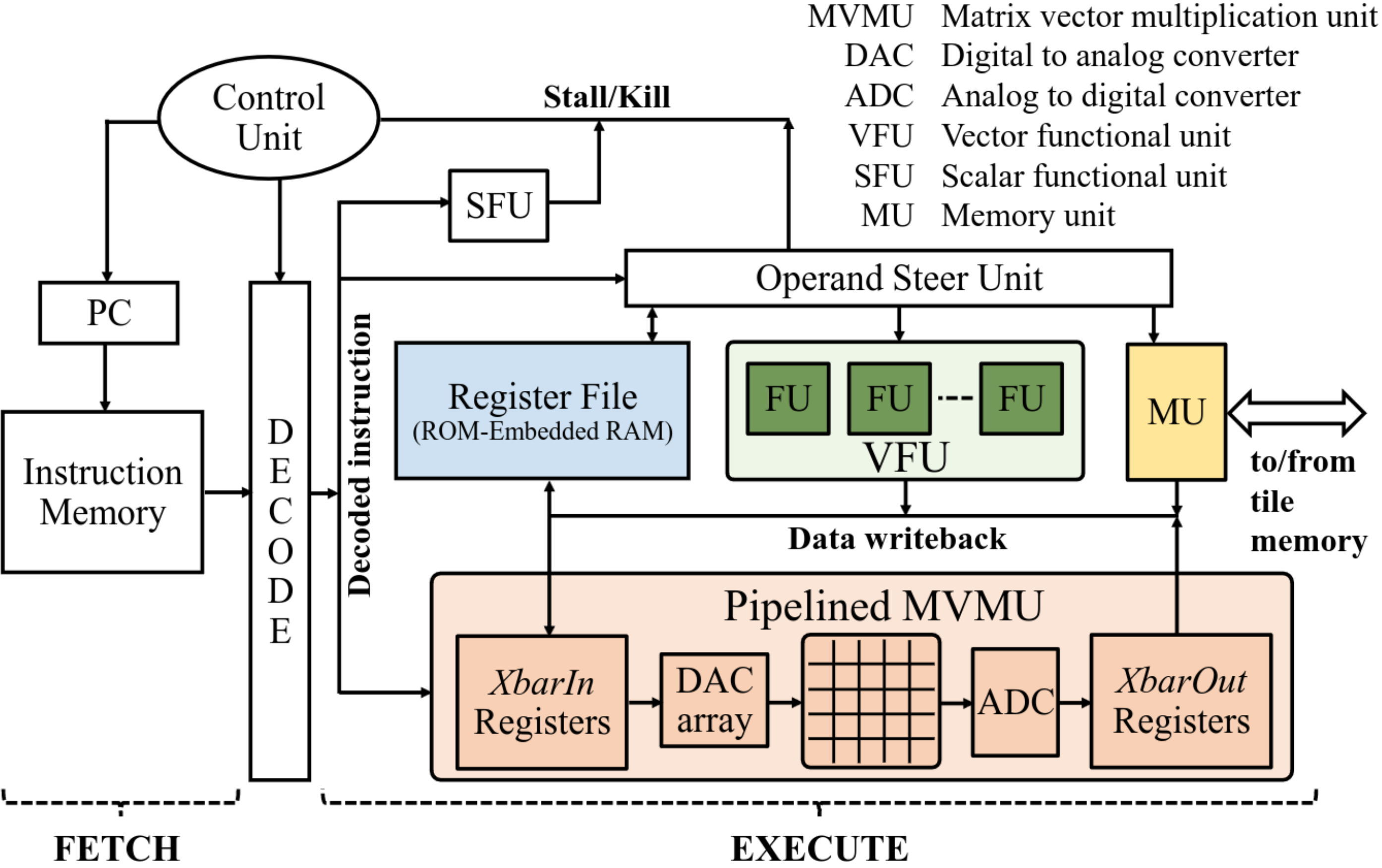}
        \end{minipage}%
    }
     \caption{PUMA tile and core architectures~\cite{puma}.}
     \label{fig:puma}
\end{figure*}

PUMA (programmable ultra-efficient memristor-based accelerator) is a generalization of memristive crossbars to accelerate a range of ML inference workloads~\cite{puma}. PUMA's microarchitecture techniques exposed via dedicated ISA ensure the efficiency of in-memory computing and analog circuitry while providing a high degree of programmability.
The architecture is organized into three hierarchy levels: cores, tiles, and nodes. Nodes are connected and communicate via a chip-to-chip network. Each individual node consists of tiles that are connected via an on-chip network, where each tile comprises cores that communicate via shared memory, as shown in Figure~\ref{fig:puma-tile}. A PUMA's core consists of its own memory and functional units, including the XBar array, referred to as the \ac{mvm} unit (MVMU), see Figure~\ref{fig:puma-core}.

Unlike most other \ac{cim} architectures, which are data parallel, PUMA is a spatial architecture where distinct instructions are executed by each core or tile. Since manually writing code for such architectures is extremely difficult, particularly when they have thousands of cores, PUMA has a runtime compiler implemented as a C++ library. The compiler takes the high-level input code and extracts a dataflow graph from it. The graph is then divided into subgraphs, considering the sizes of MVMUs, and hierarchically assigned to MVMUs, cores, and tiles. The subgraph execution is carefully scheduled to ensure effective resource utilization while avoiding potential deadlocks. Given the constraint of serial read and write operations in RRAM, PUMA exclusively supports the ML inference. However, to facilitate training, PUMA has been repurposed in a follow-up work named PANTHER~\cite{panther}.

\subsubsection{Pinatubo: Accelerating bulk bitwise logic operation}
\label{sss:pinatubo}
Pinatubo is a memristor-based architecture that harnesses data-level parallelism to conduct bulk bitwise operations~\cite{li2016pinatubo}. Unlike the crossbar configurations, it performs computations in the digital domain by modifying the \ac{sa}s. The system architecture is similar to a typical Von Neumann architecture that has a processor equipped with caches and a non-volatile main memory. Pinatubo then exploits the physical attributes of the NVM-based main memory and modifies the \ac{sa}s to support it. The main idea is the operands are stored in different rows but the same columns in an array, the rows are activated in parallel and the accumulated current in the bitline is compared to a reference level in the \ac{sa}s. For different logic gates, the memory controller changes the reference levels in the \ac{sa}s to different stats. 

\begin{figure}[tbh]
\centering
\includegraphics[scale=0.12]{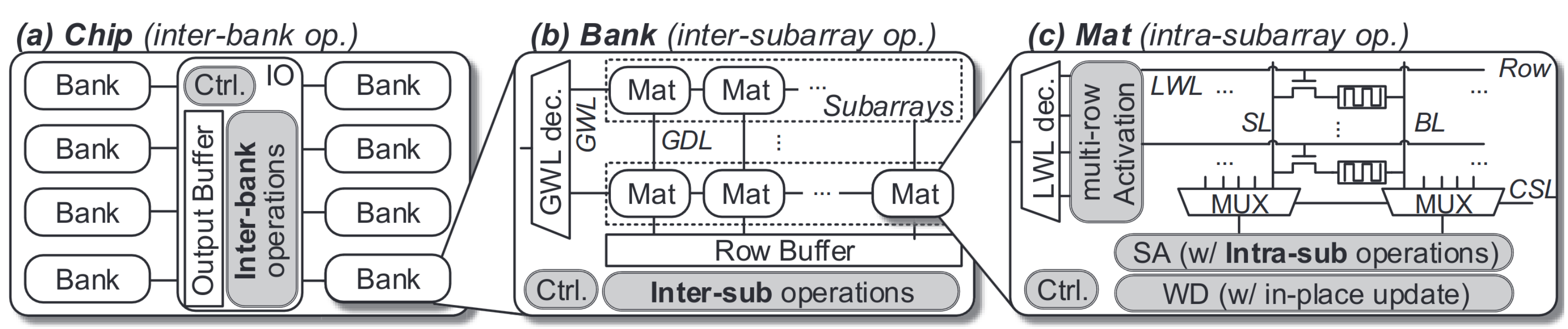}
\caption{Pinatubo architecture showing chip, bank and mat~\cite{li2016pinatubo}.} 
\label{fig:pinatubo}
\end{figure}

The Pinatubo's main memory structure is illustrated in Figure~\ref{fig:pinatubo}, comprising multiple banks divided into banks and mats. For operands within the same mat, the modified \ac{sa}s work out of the box and can perform bitwise vector operations. For operations where the data is spread across different mats, whether within the same bank or not, additional logic gates are used for execution (within the global data line or global I/O). The architecture supports only logic operations. 

For programmability, Pinatubo presents a software infrastructure containing both the programming model and runtime support components. The programming model offers two functions to allocate bit-vectors and perform bitwise operations. The runtime support facet encompasses adjustments to the C/C++ runtime library and the operating system (OS) and the development of a dynamic linked driver library. The runtime library ensures that bit-vectors are allocated to separate memory rows while the OS equipped with PIM-aware memory management, ensures intelligent invocation of the operations. 

\subsubsection{PRIME}
\label{sss:prime}
PRIME~\cite{prime}  is another RRAM-based analog \ac{cim} accelerator. The architecture comprises multiple banks where each bank integrates eight subarrays (chips) which are further (logically) split into memory (Mem) units, two full function (FF) units, and one buffer. FFs can function conventionally as memory or in an NN computation mode, controlled by the PRIME controller. A typical FF unit is 256 × 256 RRAM cells, with 6-bit reconfigurable local \ac{sa}s reading their outputs. During computation mode, RRAM resolution is 4-bit \ac{mlc}, shifting to \ac{slc} in memory mode. Distinct crossbar arrays are utilized for storing positive and negative weights. The input to the mat comes from a 3-bit fixed point signal originating from a wordline decoder and driver (WDD). Analog subtraction and sigmoid functions within the NN are implemented in the modified column multiplexers within the RRAM arrays.

\begin{figure}[tbh]
\centering
\includegraphics[scale=0.1]{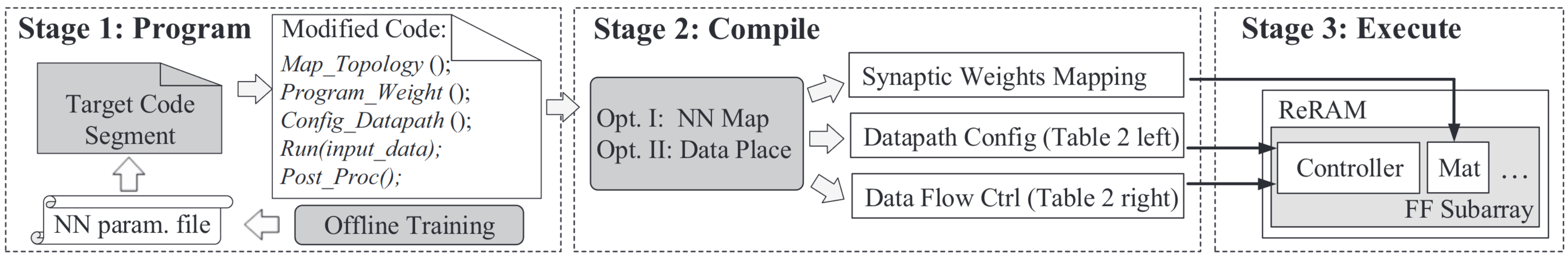}
\caption{PRIME: Source code to execution~\cite{prime}.} 
\label{fig:prime}
\end{figure}

The execution of an \ac{nn} on PRIME involves three stages. Firstly, the \ac{nn} is mapped onto FF subarrays, and synaptic weights are programmed into ReRAM cells. In the optimization stage, depending on the \ac{nn} size, mapping could occur in a single bank or across multiple banks. These first two stages are executed by the CPU. Subsequently, a series of generated instructions are transmitted to the PRIME controller in RRAM banks to perform computations. The presence of latches and OR gates facilitates pipelined computation within PRIME. 

As shown in Figure~\ref{fig:prime}, PRIME also comes with a compiler and an API, exposing device capabilities as function calls. The process from code to execution involves programming (coding), compiling (code optimization), and code execution. PRIME offers application programming interfaces (APIs) that empower developers to map \ac{nn} topologies onto FFs and configure data paths etc.

\subsubsection{Pipelayer}
\label{sss:pipelayer}
\begin{figure}[tbh]
\centering
\includegraphics[scale=0.12]{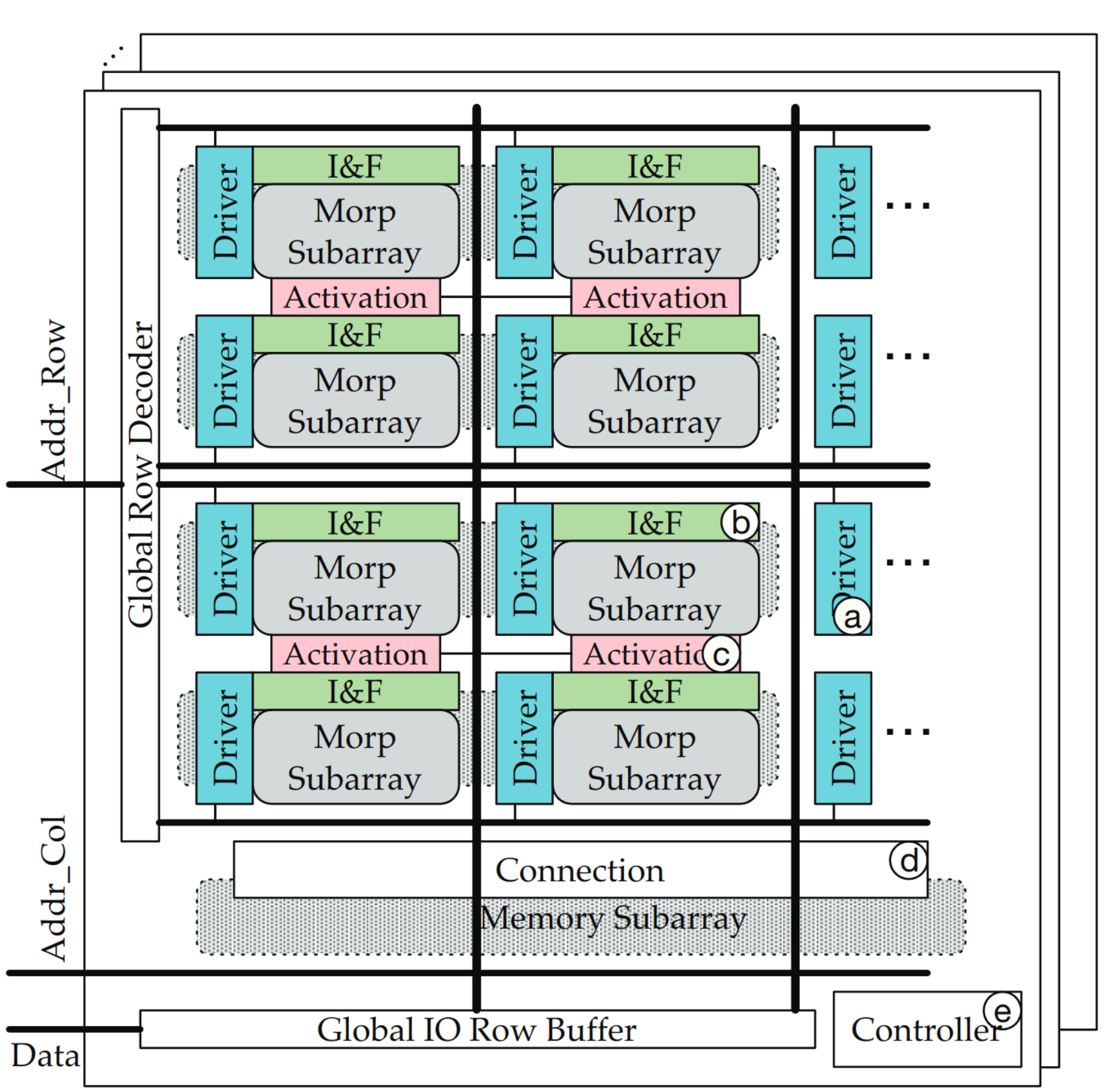}
\caption{An overview of the Pipelayer architecture~\cite{pipelayer}.} 
\label{fig:pipelayer}
\end{figure}

Pipelayer is another RRAM-based accelerator for C\ac{nn}s that supports both training and inference~\cite{pipelayer}. The overall architecture of PipeLayer, shown in Figure~\ref{fig:pipelayer} features RRAM crossbars, the spike Driver block to encode inputs as spikes and get rid of DACs, and integration and fire components that eliminate ADCs. In write mode, the spike driver updates RRAM array weights with a 4-bit resolution. Within the cell, data processing occurs across morphable and memory subarrays, where memory subarrays are conventional memory arrays while morphable arrays can be configured in both compute and memory modes. Pipelayer leverages these morphable subarrays for different purposes in training and inference. 

PipeLayer allows interactive configuration of the system on a per-layer basis. It provides an API that has functions for different operations e.g.,  bidirectional transfer of data between the CPU main memory and PipeLayer, the topology\_set function, where the number of compute groups can be specified by the programmer, the weight\_load function to load either pre-trained weights during testing or initial weights during training into the arrays. Other functions include pipeline and mode set functions for the morphable subarrays.

There are many other RRAM-based analog and digital \ac{cim} accelerators. Some other common ones that are mostly taped-out and not discussed here include: AtomLayer~\cite{atomlayer}, RIMAC\cite{rimac}, FORM~\cite{forms}, RRAMs for pattern recognition ~\cite{wang2018fully}, RRAM accelerator for BNNs (ISSCC, 65 nm)~\cite{chen201865nm}, RRAM for edge processors (ISSCC, 55 nm)~\cite{xue201924}, analog RRAM with fully parallel MAC and extremely high TOPS/W (ISSCC, 130 nm but large array)~\cite{liu202033}.

\subsubsection{In-DRAM computing}
\label{sss:ambit}
Ambit~\cite{ambit} is a DRAM-based \ac{cim} accelerator that, unlike all previous systems, leverages the analog capabilities of current DRAM technology for executing bulk bitwise operations. Ambit mainly comprises two components. First, Ambit-AND-OR implements \ac{tra} in conventional DRAM. Like memristors, the idea is to activate three rows in parallel and leverage the \emph{charge-sharing and charge accumulation principle}. \ac{tra} produces a bitwise majority function. Controlling the initial state of one of the three rows enable performing AND and OR operation. The second component of Ambit is Ambit-NOT, which uses the inverters in the DRAM \ac{sa}s to implement the logic NOT operation. The basic components are then extended to implement other logic operations and accelerate bulk bitwise operations in multiple applications. 
With 8 DRAM banks, Ambit demonstrates a substantial improvement in bulk bitwise operation throughput compared to an Intel Skylake processor and the NVIDIA GTX 745 GPU.

A follow-up work on the bulk bitwise logic in DRAM, ComputeDRAM~\cite{computedram} demonstrated that by deliberately violating timing parameters between activation commands, certain existing off-the-shelf DRAM chips can implement the \ac{tra} operation of Ambit. This indicates that certain real-world off-the-shelf DRAM chips, despite not being intended for Ambit operations, can indeed perform in-DRAM AND and OR operations. This also suggests that the concepts introduced in Ambit might not be too far from practical implementation. If existing DRAM chips can perform such operations to some extent, then chips explicitly designed for such functions could potentially be even more capable. 

\subsubsection{In-SRAM computing}
\label{sss:imac}
Neural Cache~\cite{neuralcache} is an SRAM-based \ac{cim} accelerator primarily targeting \ac{cnn}s. The core operations of Neural Cache are bitwise AND and NOR operations, which are executed by simultaneously activating multiple rows (charge sharing). It repurposes the cache memory by modifying the peripheral circuitry to support operations such as convolution, pooling, quantization, and fully-connected layers, all performed at an 8-bit data precision. It is also capable of performing bit-serial operations like addition, subtraction, multiplication, comparison, search, and copy for larger data, utilizing carry latches linked to \ac{sa}s. A transpose memory unit is introduced that facilitates the reorganization of data into bit-serial format within the memory when needed.

IMAC~\cite{imac} is another SRAM-based \ac{cim} accelerator that uses the precharge circuit to perform multi-bit analog multiplication by encoding the bit significance in the pulse width of pre-charge pulse. IMAC also requires DAC/ADC converters to facilitate the conversion between digital and analog forms. 
There are many other instances of SRAM-based \ac{cim} designs, some even \textbf{taped-out}~\cite{sram-cim-chip, valavi201964, conv-SRAM, xnor-SRAM}.

\subsubsection{In-MRAM computing}
\label{sss:cim-mram}
In NVMs, Magnetic RAM (MRAM) is probably the most mature memory technology that is commercially available and is already used in many embedded devices (see Section~\ref{subsec:mram}). Therefore, it has also been intensively investigated in the \ac{cim} context and computing approaches implementing in-MRAM basic boolean logic operations and more complex arithmetic functions have been showcased. Like all other technologies, the basic \ac{cim} methods include bit-cell modification, reference adaptation, and in-memory analog computation.

\begin{figure}[tbh]
\centering
\includegraphics[scale=0.10]{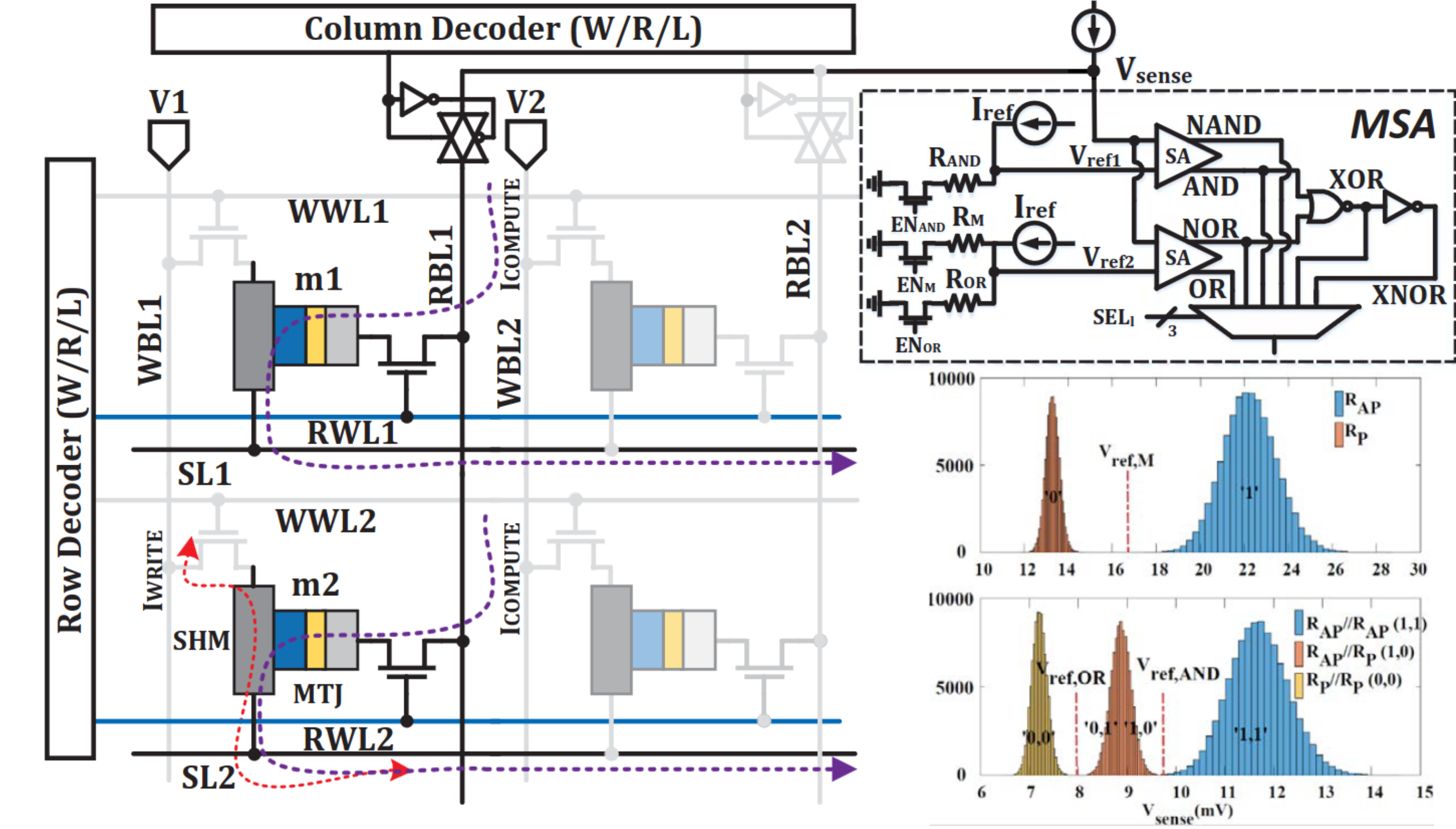}
\caption{A typical SOT-MRAM subarray architecture for in-place logic~\cite{cmp-pim}.} 
\label{fig:cim-MRAM}
\end{figure}

Figure~\ref{fig:cim-MRAM} shows a typical subarray architecture of an in-MRAM \ac{cim}~\cite{cmp-pim}. The important difference here compared to already known aspects is that it has separate read and write bit and word lines and three reference resistance states in the sensing circuity (RAND/RM/ROR). RM is used to perform normal memory operations, while RAND and ROR, as the names suggest, are used to implement AND and OR operations, respectively. 

Similar to other technologies, the boolean logic is implemented with charge-sharing, and MAC is implemented in the analog domain with current accumulation. Some prominent MRAM-based \ac{cim} designs include analog MACs for \ac{tra} inference~\cite{analog-cim-MRAM}, MRAM-\ac{cim} for BNNs~\cite{chang2019pxnor}, and MRAM crossbar~\cite{samsung-mram-crossbar}.

\subsubsection{CIM using FeFETs}
\label{sss:cim-fefet}
FeFeTs have also been shown to implement in-place logic gates, addition, and content-addressable memories (CAMs). Notably, these logic operations can also be implemented with a single \ac{fefet} cell. For instance, if one operand is stored in a cell (or a set of cells), the other operand can be applied as input to perform logic operation~\cite{cim-fefet-namlab}, akin to the working principle of crossbars. Further, solutions proposing activating multiple rows and leveraging the bitline's charge sharing (as in other memory technology) have also been presented~\cite{cim-fefet}.

FeFETs have received particular interest in CAM designs. CAMs are associative memories that can perform parallel searches for a query across all stored contents within an array. FeFeTs have been used to implement different types of CAMs for exact-search operations, approximate search operations, range-based search operations, or a combination of them~\cite{fefet-cam-universal, fefet-approx-search}. 

\subsubsection{Latest industrial chips}
\label{sss:cim-latest}
In the previous sections, we have extensively discussed a variety of notable \ac{cim} and \ac{cnm} solutions employing different technologies. While a few of these systems have been developed in collaboration with industry partners and a subset has undergone the tape-out process, the majority of these accelerators originate from academia. In this section, we specifically present \ac{cim} systems originating from the industrial sector in the last couple of years. Note that these \ac{cim} systems also primarily show prototypes showcasing various research outcomes, but they indicate their potential realization in the near future.

~\\
\noindent\textbf{IBM's PCM-based accelerators:}
For more than five years, IBM has been using its PCM device to do in-place operations for different use cases. Initially, they were working with a reservoir of devices (millions of them) and implementing the peripheral circuitry and additional CMOS logic in an FPGA. Their research has progressed to consolidate all components onto a single chip, as exemplified by HERMES, a core composed of 256$\times$256 PCM array with ADCs, a local digital processing unit, and additional peripheries~\cite{IBM-hermes}. 
The core effectively executes a fully parallel 256×256 analog \ac{mvm}, where each 8T4R unit cell encodes a positive/negative weight, with simultaneous handling of 256 8-bit digital inputs/outputs. 
Positive weights are encoded by combining the conductance of two PCM devices, while negative weights are represented by the other two PCMs within the unit cell. 


This year, IBM announced a newer 64-core \ac{cim} chip designed and fabricated in 14-nm CMOS technology integrated with PCM~\cite{ibm-64-cores}. The fully integrated chip comprises 64 cores, each with a size of 256$\times$256, connected through an on-chip communication network. It reportedly achieves an unparalleled maximal throughput of 63.1 TOPS at an energy efficiency of 9.76 TOPS/W for 8-bit input/output \ac{mvm}s. 

~\\
\noindent\textbf{Samsung's MRAM crossbar:}
Crossbar-based analog \ac{mvm} is well-explored in RRAM and PCM technologies.
However, implementing MRAM-based crossbars is challenging due to the inherent low resistance of these devices, which could lead to significant power consumption. In 2022, Samsung presented a 64x64 MRAM crossbar array to address the low-resistance issue by employing an architecture that uses resistance summation (instead of current summation) for analog multiply-accumulate operations~\cite{samsung-mram-crossbar}. 
Compared to the IBM HERMES cores, Samsung's crossbar is significantly less sophisticates and limited in scale.

~\\
\noindent\textbf{TSMC's in-SRAM accelerator:}
While other SRAM-based \ac{cim} chips exist, our focus is on the TSMC macro structure using standard 8T cells~\cite{tsmc-SRAM} due to its better noise margin, ensuring stable activation for multiple rows operations in the \ac{cim} mode, albeit with approximately 30\% increased area. 

\begin{figure}[tbh]
\centering
\includegraphics[scale=0.10]{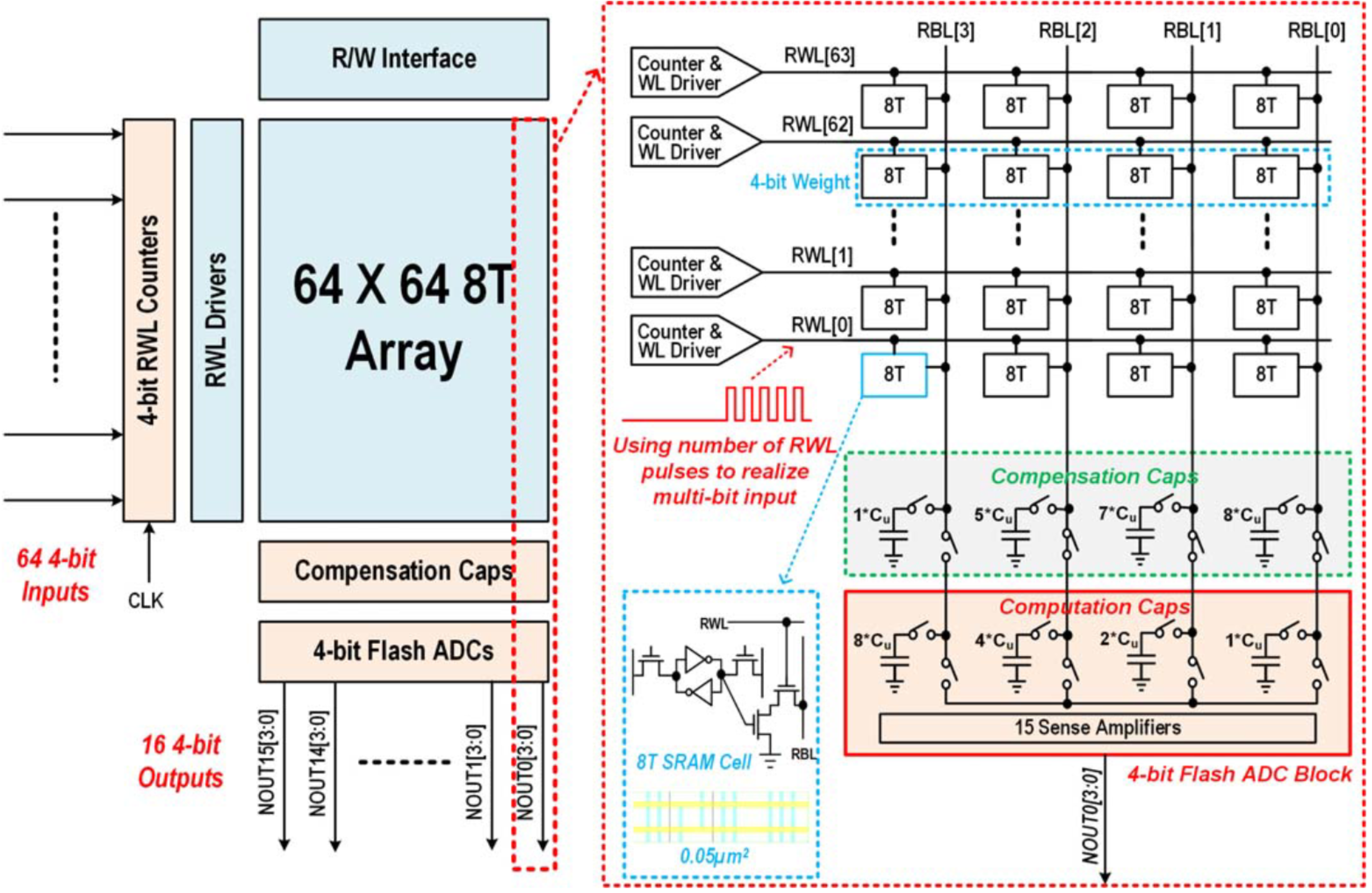}
\caption{TSMC's \ac{cim} SRAM structure~\cite{tsmc-SRAM}.} 
\label{fig:tsmc-cim-sram}
\end{figure}

The proposed design shown in Figure~\ref{fig:tsmc-cim-sram} has a 64$\times$64 SRAM array and enables parallel computations of the multiply-and-average (MAV) operations. In a single cycle, the MAV computation of 64 4-bit inputs with 16 4-bit weight can be completed. The 4-bit input is represented by the number of read word line pulses which is precisely controlled by 4-bit digital counters. The 4-bit weight is achieved through charge sharing across binary-weighted computation capacitors. Each computation capacitor unit is constructed using the inherent capacitor of the \ac{sa} within the 4-bit flash ADC to optimize space and minimize the kick-back effect. 
This 64x64 8T macro is fabricated using 7nm FinFET technology, exhibiting an energy efficiency of 351 TOPS/W and a throughput of 372.4 GOPS for 1024 (64x16) 4x4b MAV operations.

\begin{figure}[tbh]
\centering
\includegraphics[scale=0.1]{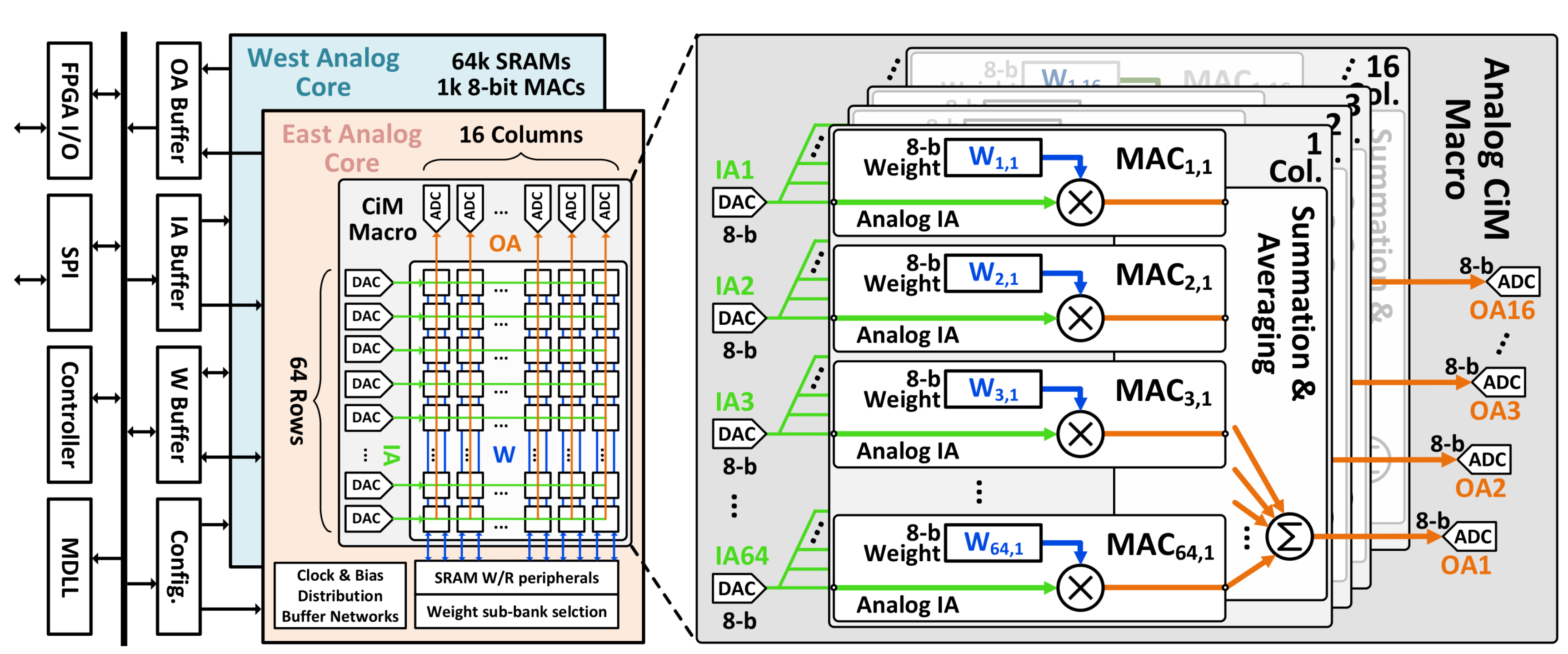}
\caption{Chip level architecture diagram of Intel's analog \ac{cim} design~\cite{intel-sram-cim}.} 
\label{fig:intel-cim}
\end{figure}

~\\
\noindent\textbf{Intel's SRAM-based analog CIM design:}
Intel has recently proposed an SRAM-based \ac{cim} macro utilizing their 22nm Low-Power FinFET process~\cite{intel-sram-cim}. Through the implementation of a 1-to-2 ratioed capacitor ladder (C-2C)-based charge domain computing scheme, the presented prototype chip (shown in Figure~\ref{fig:intel-cim}) achieves the capability to perform up to 2k MAC operations in a single clock cycle, alongside achieving a peak power efficiency of 32.2-TOPS/W with 8-bit precision for both input activation and weights. The chip also ensures accurate \ac{mvm}s by restricting the computation error of less than 0.5\%.

~\\
\noindent\textbf{Bosch+Fraunhofer and GlobalFoundries+Fraunhofer FeFET based CIM designs:}
Fraunhofer is also actively working on exploring the manufacturability and scalability aspects of FeFET and MRAM devices at both the device and array levels. Together with GlobalFoundries, they have demonstrated a FeFET-based crossbar array for multiply-accumulate (MAC) operation~\cite{gf-fefet-crossbar}. The array was fabricated at GlobalFoundries with 28nm CMOS technology coupled with FeFET. To prevent the accumulation of errors on the bitline, the arrays were divided into 8$\times$8 segments.

In a recent work, Fraunhofer and Robert Bosch demonstrated a \ac{cim} crossbar using multi-level FeFET cells. In the proposed design, the input is encoded into the applied voltage duration and magnitude while the weights are stored in the multi-level FeFET cells. The MAC output is the accumulated capacitor voltage that depends on the activation time and the number of FeFETs activated. This reportedly reduces the impact of variations and the achieved performance of 885.4 TOPS/W is also nearly-double compared to existing solutions.  

\begin{figure}[tbh]
\centering
\includegraphics[scale=0.18]{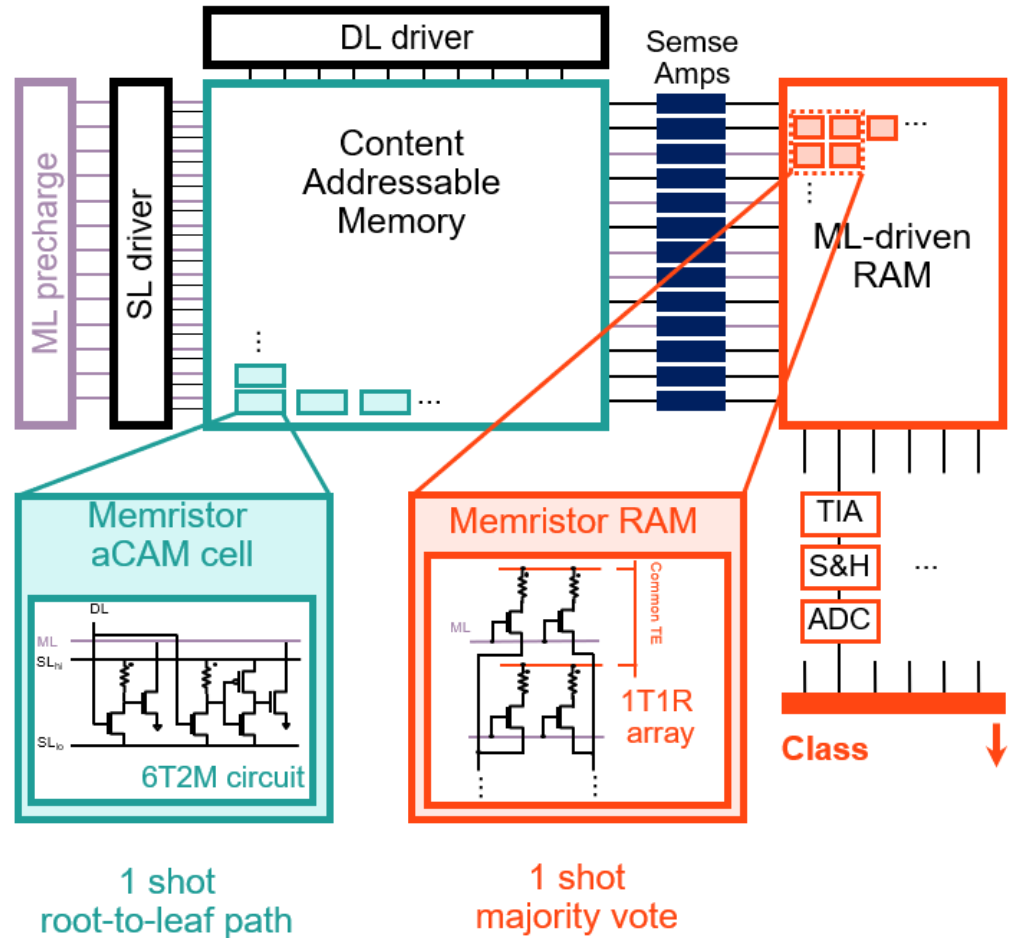}
\caption{An overview of the HP \ac{cim} system for tree-based learning~\cite{hp-cams}.} 
\label{fig:hp-cams}
\end{figure}

~\\
\noindent\textbf{HP's CAM designs:}
In a recent work, Hewlett Packard Labs proposed a memristive-based analog CAM for tree-based machine learning~\cite{hp-cams}. Analog CAMs are capable of performing searches based on analog signal levels rather than digital data comparison. The proposed design combines analog CAMs with traditional analog RAM and accelerates large random forest models with it. Figure~\ref{fig:hp-cams} shows a high-level overview of the proposed system where the analog CAM can perform root-to-leaf evaluation of an entire tree in a single step. 

\subsection{Comparative analysis and discussion}
\label{subsec:compare}
The surge in \ac{cim} and \ac{cnm} systems is largely attributed to the revolution in data-intensive applications. According to recent research by TSMC, traditional SRAM and DRAM technologies have effectively scaled to meet capacity and bandwidth demands in the past decades~\cite{datagrowth}, but their future scalability is uncertain due to reaching inherent technological limits. This underscores the pivotal role that NVM will play in the future of computing.

Especially in edge scenarios such as automotive, augmented reality, and AI, where energy efficiency is paramount, NVM technologies are poised to play a pivotal role. As energy efficiency increases through specialized hardware, domain-specific architectures harnessing these NVMs for \ac{cim} and \ac{cnm} solutions are anticipated to experience an unprecedented surge in the coming years.
A recent article from Intel~\cite{intel-sram-cim} compares the performance of conventional digital accelerators with the emerging analog and digital \ac{cim} and \ac{cnm} accelerators. Conventional accelerators still achieve higher throughput because \ac{cim} systems are relatively less optimized, array sizes are small, and the peripheral circuitry overhead is non-negligible. Yet, they are orders of magnitude better in terms of power consumption. As of the time of writing, the most recent comparison depicted in ~\cite{datagrowth} shows similar trends.

\begin{figure}[tbh]
\centering
\includegraphics[scale=0.2]{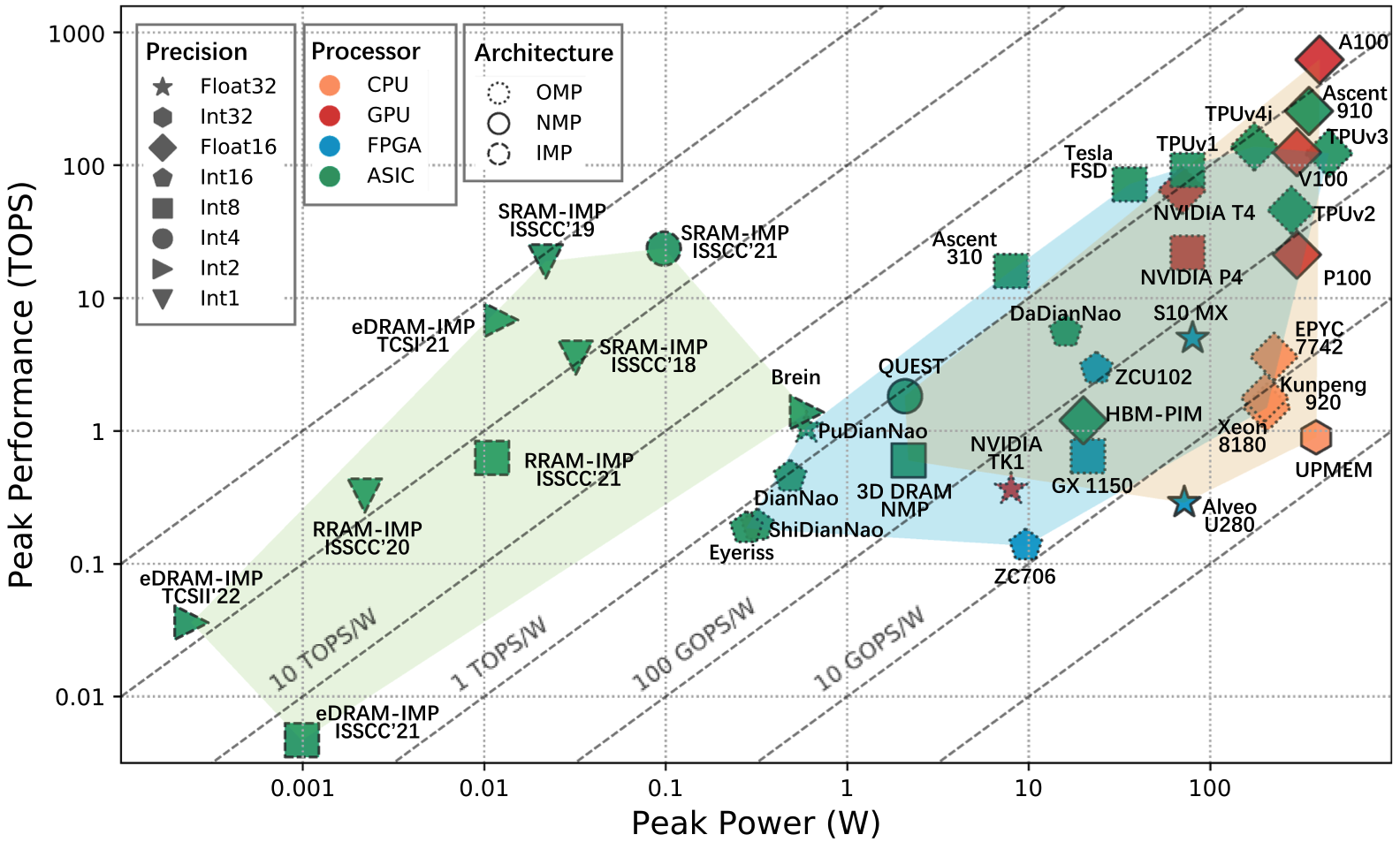}
\caption{Performance and power comparison of different \ac{omp} (we call it \ac{com} in this report), \ac{nmp} (ours \ac{cnm}) and \ac{imp} (ours \ac{cim})~\cite{zhang2023survey}.} 
\label{fig:cim-cnm-conv-comp}
\end{figure}

Table~\ref{fig:arch-comp} presents a summary and comparison of the architectures discussed in this section. For brevity, we only compare important parameters, such as the underlying memory technology, available function (boolean logic, arithmetic, etc.), evaluation technique (simulation, prototype, analytic), programming model, application domain, and technology node.

\begin{table}
    \scriptsize
        \begin{center}
        \begin{tabular}{p{0.10\linewidth}p{0.03\linewidth}p{0.1\linewidth}p{0.03\linewidth}
        p{0.12\linewidth}p{0.12\linewidth}p{0.13\linewidth}p{0.15\linewidth}}
        \toprule
        \textbf{Accelerator} & \textbf{Year} & \textbf{Technology} & \textbf{Type} & \textbf{Programming model} & \textbf{Logic unit} & \textbf{Implementation} & \textbf{Domain} \\
        \midrule
        McDRAM & 2018 & DRAM & CNM & Extended ISA & MAC & Hardware & AI \\
        MViD & 2020 & DRAM & CNM & Extended ISA & MAC & Hardware & AI \\
        PIM-HBM & 2021 & DRAM (HBM) & CNM & Full software-stack & FPUs (add, Mul) & Hardware & AI \\ 
        AiM & 2022 & DRAM (GDDR6) & CNM & API & MAC & Hardware & AI \\ 
        AxRAM & 2018 & DRAM (GPU-based system) & CNM & API & MAC, LUTs & GPGPU-Sim & AI \\ 
        TESSERACT & 2015 & DRAM (HMC) & CNM & API & CPU & Simulation & Graph processing \\
        TOP-PIM  & 2014 & DRAM (HMC) & CNM & OpenCL & CPU+GPU & Simulation & Graph, HPC \\ 
        AMC & 2015 & DRAM (HMC) & CNM & OpenMP & CPU & Simulation & HPC \\ 
        HRL & 2015 & DRAM (HMC) & CNM & MapReduce & CGRA+FPGA & Simulation & Data analytics \\
        \midrule
       \multicolumn{8}{c}{\textbf{CIM architectures (Academia/Research Labs)}}  \\
       \midrule
       ISAAC & 2016 & RRAM & CIM & NA & Analog Xbar & Analytical & AI \\
       PUMA & 2019 & RRAM & CIM & Compiler & Xbar & PUMAsim (arch. simulation) & AI \\
       Pinatubo & 2016 & RRAM & CIM & API, Runtime & Boolean logic & In-house simulator & Bitwise Logic \\
       PRIME & 2016 & RRAM & CIM & Compiler+API & Xbar & Analytical & AI \\
       PipeLayer & 2017 & RRAM & CIM & API & Xbar & Analytical & CNN (train + infer)\\
       AtomLayer & 2018 & RRAM & CIM & NA & Xbar & Analytical & CNN (train + infer)\\
       RIMAC & 2023 & RRAM & CIM & NA & Xbar (without DAC/ADC) & In-house simulator & DNN inference  \\    
       
       \bottomrule
\end{tabular}
\caption{A summary of the presented architectures. They are grouped into three categories: \ac{cnm}, \ac{cim}, and \ac{cim} (prototype chips/systems). All presented architectures are either simulation-based or prototype-based (no products).} 
\label{fig:arch-comp}
\end{center}

\end{table}

%% file: contents/cim-startups.tex
This section overviews CIM and CNM companies/startups, highlighting their products, underlying technologies, customers (when known), and tools. As not everything about companies is public, we only include details that we extract from these companies' websites or are known to us via our network. 

\subsection{Axelera}
\label{subsec:axelera}
Axelera~\cite{axelera} is one of the notable Semiconductor startups in Europe. Founded in 2021 and backed by tech giants like Bitfury and IMEC, it had already taped out its first CIM chip, Thetis, in December 2021 (just four months after its founding). Today, it offers a fully integrated \ac{soc} powered by its Metis AI processing units (AIPU).   

About the AI core, as per the company's website: ``Axelera AI has fundamentally changed the architecture of “compute-in-place” by introducing an SRAM-based digital in-memory computing (D-IMC) engine. In contrast to analog in-memory computing approaches, Axelera’s D-IMC design is immune to noise and memory non-idealities that affect the precision of the analog matrix-vector operations as well as the deterministic nature and repeatability of the matrix-vector multiplication results. Our D-IMC supports INT8 activations and weights, but the accumulation maintains full precision at INT32, which enables state-of-the-art FP32 iso-accuracy for a wide range of applications without the need for retraining''. 

Axelera's latest SoC consists of 4 cores and a RISC-V based control core. For programming these systems, Axelera provides an end-to-end \underline{integrated framework} for application development. The high-level framework takes users along the development processes without needing to understand the underlying architecture or even the machine learning concepts. 

\vspace{0.1cm}\noindent\textbf{Funding:} ``Axelera AI, the provider of the world’s most powerful and advanced solutions for AI at the Edge, announces new investors who have joined their oversubscribed Series A round, bringing the total amount raised to \$50 million. In the last several months, CDP Venture Capital, Verve Ventures, and Fractionelera have joined the round'', Axelera AI, May 22, 2023.

\subsection{d-Matrix}
\label{subsec:d-matrix}
d-Matrix is at the forefront of driving the transformation in data center architecture toward digital in-memory computing (DIMC)~\cite{d-matrix}. Founded in 2019, the company has received substantial support from prominent investors and strategic partners, including Playground Global, M12 (Microsoft Venture Fund), SK Hynix, Nautilus Venture Partners, Marvell Technology, and Entrada Ventures.

Leveraging their in-SRAM digital computing techniques, a chipset-based design, high-bandwidth BoW interconnects, and a full stack of machine learning and large language model tools and software, d-Matrix pioneers best-performing solutions for large-scale inference requirements.
 A \underline{full stack framework}, compiler, and APIs (open-source as per the company's website but couldn't find the link).
Their latest product Jayhawk II can scale up to 150 TOPS/W using 6nm technology and can handle LLM models up to 20$\times$ more inferences per second for LLM sizing to 40B parameters, compared to state-of-the-art GPUs. 

\vspace{0.1cm}\noindent\textbf{Funding:} Temasek, Playground Global and Microsoft Corp.

\subsection{Gyrfalcon Technology}
\label{subsec:gyrofalcon}
Gyrfalcon Technology~\cite{gyrofalcon} also leverages CNM to accelerate AI on the edge. They offer an AI processing in memory (APiM) architecture that combines a large MAC array directly with MRAM memory modules. As of the current date, their \underline{software stack} is not available.

\vspace{0.1cm}\noindent\textbf{Funding:} Private.

\subsection{MemComputing}
\label{subsec:memcpu}
MemComputing~\cite{memcpu}, founded in 2016, uses a computational memory based on its self-organizing logic gates (SOLG). SOLGs are terminal-agnostic elements (memristor or memcapacitor) that implement various logic gates. Their target applications comprise industrial computations associated with optimizations, big data analytics, and machine learning. MemComputing provides a \underline{software stack} and offers it as a software-as-a-service. 

\vspace{0.1cm}\noindent\textbf{Funding:} MemComputing mentions the US Space Force, ENSOS, NASA, Ball Aerospace, PSA, US Air Force, Canvass Labs and Defence Innovation Unit as partners.

\subsection{Memverge}
\label{subsec:memverge}
Memverge~\cite{memverge} is not directly doing any CIM or CNM but is relevant in the context. Backed by 9 investors including tech giants like Intel, SK hynix, the company's main goal is to provide software designed to accelerate and optimize data-intensive applications. Their main target is to consider environments with ``Endless Memory'' and efficiently manage the memory to get more performance. 

\vspace{0.1cm}\noindent\textbf{Latest news}: ``Samsung, MemVerge, H3 Platform, and XConn, today unveiled a 2TB Pooled CXL Memory System at Flash Memory Summit. The system addresses performance challenges faced by highly distributed AI/ML applications. These challenges include issues like spilling memory to slow storage when main memory is full, excessive memory copying, I/O to storage, serialization/deserialization, and Out-of-Memory errors that can crash an application.'', MemVerge, August 8, 2023.

\subsection{Mythic}
\label{subsec:mythic}
\begin{figure}[tbh]
\centering
\includegraphics[scale=0.35]{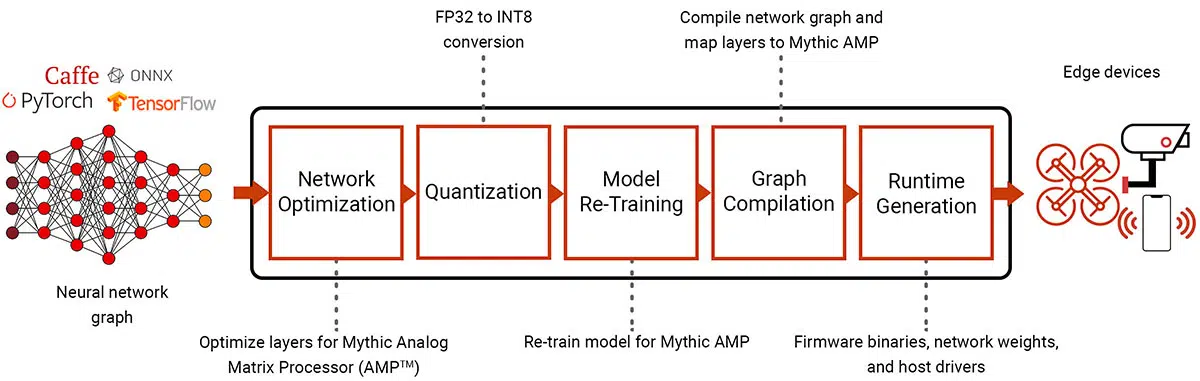}
\caption{Mythic AI workflow~\cite{mythic}.} 
\label{fig:mythic-workflow}
\end{figure}

Mythic~\cite{mythic} offers an analog matrix processor (Mythic AMP) that uses their analog compute engine (ACE) based on flash memory array and ADCs. Mythic ACE also has a 32b RISC V processor, SIMD vector engine, and a 64KB SRAM along with a high-throughput network-on-chip (NoC).
Mythic workflow in Figure~\ref{fig:mythic-workflow} shows that the \underline{software stack} takes a trained NN model, optimizes it, and compiles it to generate code for Mythic AMP. The optimization suit also transforms NN in a way that can be accelerated on the analog CIM system. 

\vspace{0.1cm}\noindent\textbf{Funding:} The company is supported by many investors: Micron, HP Enterprise, SoftBank, Future ventures, Lam Research, Threshold, Catapult, DCVC and UDC ventures. 

\subsection{NeuroBlad}
\label{subsec:neuroblad}
Founded in 2018, NeuroBlade offers the SPU (SQL Processing Unit), the industry’s first, proven processor architecture that delivers orders of magnitude improvement by departing from Von Neumann model~\cite{neuroblad}. Neuroblade is also a CNM architecture (more closed to near-storage computing) where they integrate custom RISC processors on the DRAM chip (very similar to UPMEM). The SPUs are installed as PCI-e cards that can be deployed in data centers. As for the \underline{software stack}, the company offers an SDK along with a set of APIs that hide the complexity and programming model for these cores from the end user and also allow optimizing for maximum parallelism and efficiency.

\vspace{0.1cm}\noindent\textbf{Funding:} NeuroBlade is funded by Stage one, Grove Ventures, UMC, PSMC Intel capitals, Pegratron, Marubeni, Marius Nacht, Corner and MediaTek. 

\subsection{Rain AI}
\label{subsec:rain}
Founded in 2017, Rain AI also focuses on radically cheaper AI computing~\cite{rainai}. The company has no hardware product yet but is aiming to be 100$\times$ better than GPU using their innovations in radical co-design (by looking at the algorithms and the CIM hardware at the same time). They are targeting AI training (along with the inference) on the edge with the ultimate goal of putting models the size of ChatGPT into chips of the size of a thumbnail. They are transforming the algorithms in a way that fundamentally matches the behavior of the analog memristive devices. As per the CEO, they have a few tap-outs planned for this year and the product (a complete platform) next year and they are working on a \underline{software stack} for ease of use and ease of integration. 

\vspace{0.1cm}\noindent\textbf{Funding:} The company is funded by Y combinator S18, Sam Altman (CEO OpenAI), Liquid 2 Ventures, Loup Ventures, Airbus Ventures, and Daniel Gross (founder Poineer). 
\subsection{SEMRON}
\label{subsec:semron}
Founded in 2020, Semron~\cite{semron} promises to offer 3D solutions powered by analog CIM. At the core of their technology is their innovative CapRAM devices which are semiconductor devices that store multi-bit values in their variable capacitances (unlike variable resistance states in memristors). 
Since CapRAM is capacitive, the noise in calculations is much lower and the energy efficiency, as per their website, is unparalleled. 
Although Semron has the device technology, there are no details of its products, architecture, and \underline{software stack}. 

\vspace{0.1cm}\noindent\textbf{Funding:} As per crunchbase, the company is funded by VentureOut. 
\subsection{SureCore}
\label{subsec:surecore}
Surecore~\cite{surecore} is working on many low-power products including custom application-specific. They also have a product named ``CompuRAM'' that embeds arithmetic capability within the SRAM array to enable low-power AI on the edge. 
Besides working on SRAM-based solutions, in collaboration with Intrinsic, they have recently ventured into RRAM technology.
No information is provided regarding the \underline{software stack}.

\vspace{0.1cm}\noindent\textbf{Funding:} The company is supported by Capital-E, Finance Yorkshire and Mercia Technologies. 
\subsection{Synthara}
\label{subsec:synthara}
Synthara is a Zurich-based Semiconductor company that was founded in 2017~\cite{synthara}. Their latest product, ComputeRAM, integrates SRAM-based CIM macros with proprietary elements to accelerate dot products. The solution delivers 50$\times$ compute efficiency and can be used for AI, digital signal processing, and linear algebra-heavy routines. The CIM-powered SRAM array can be operated just like conventional SRAM. ComputeRAM is not married to a specific ISA and can work with any host processor.
Synthara also provides what they call Compiler hooks that can transparently offload any input application to their ComputeRAM accelerator, without changing or rewriting the code. 

\vspace{0.1cm}\noindent\textbf{Funding:} The company is supported by EU funding for research \& innovation, High-tech Gr\"underfonds, Intel.ignite, FNSNF, multicoreware, ventureKick and others. 
\subsection{Syntiant}
\label{subsec:syntiant}
Founded in 2017, Syntiant also leverages DRAM-based \ac{cnm} and utilizes standard CMOS processes to design their neural decision processors (NDPs) that perform direct processing of neural network layers from platforms like TensorFlow~\cite{syntiant}. Syntiant also mainly targets AI on the edge having applications in many domains, including always-on voice, audio, image, and sensor applications. 

Syntiant's TinyML platform, powered by NDP101, aspires to democratize AI by presenting a comprehensive system for those interested in initiating their own model training for edge computing.

\vspace{0.1cm}\noindent\textbf{Funding:} Syntiant is funded by prominent investors including, Atlantic Bridge, Rober Bosch Venture Capital, Embark Ventures, DHVC, Intel capitals, M12 (Microsoft ventures), and Motorola Solutions.
\subsection{TetraMem}
\label{subsec:tetramem}
Founded in 2018, TetraMem is set to offer the industry's most disruptive CIM technology for edge application~\cite{tetramem}. TetraMem is also leveraging memristors for analog MAC operations, aiming at inference on the edge. Their systems are built upon their patented devices and co-design solutions.

TetraMem offers (1) Platform as a service (PaaS), a complete hardware and software platform designed to integrate into your own system; (2) Software as a service (SaaS), to help develop your NN edge application and integrate it into your system. Their verified full software stack provides an unmatched experience on actual analog in-memory compute silicon; and (3) a neural processing unit (NPU) based on memristive technology.

TetraMem has recently announced a collaboration with Andes Technologies and together with their research collaborators have demonstrated a memristive device that can have thousands of conductance levels (unmatched)~\cite{tetramem-thousands-stats}. 

\vspace{0.1cm}\noindent\textbf{Funding:} Private.
\subsection{EnCharge AI}
\label{subsec:encharge}
Founded in 2022, EnCharge AI promise to offer an end-to-end scalable architecture for AI inference~\cite{enchargeAI}. They leverage SRAM-based CIM arrays for analog MVM operations and combine them with SIMD CNM logic to perform custom element-wise operations. The architecture comprises an array of CIM units (CIMUs), an on-chip network interconnecting CIMUs, buffers, control circuitry, and off-chip interfaces. Each CIMU is equipped with an SRAM-based CIM array featuring ADCs to convert computed outputs into digital values. Additionally, CIMUs house SIMD units and FPUs with a custom instruction set, along with buffers dedicated to both computation and data flow.
According to the company's official website, they offer a \underline{software platform} that fits with standard ML frameworks, such as PyTorch, TensorFlow, and ONNX. This also allows the implementation of various ML models and their customizations. Specific implementation details about the software stack are not available.

\vspace{0.1cm}\noindent\textbf{Funding:} Encharge AI is funded by AlleyCorp, Scout Ventures, Silicon Catalyst Angels, Schams Ventures, E14 Fund, and Alumni Ventures. At their launch in December 2022, they announced securing \$21.7 Mio. in their series A round.
\subsection{Re(conceive) AI}
\label{subsec:reconceive}
Re(conceive) is another CIM startup founded in 2019 that promises offering ``the most power AI accelerator''~\cite{reconceive}. As per their website, re(conceive) are pioneers in realizing the complete potential of CMOS-based analog in-memory AI computing, achieving the utmost efficiency among all known AI accelerators. However, no specific details are available on the company's funding and \underline{technology (hardware/software)}.

\subsection{Fractile AI}
\label{subsec:fractile}
Established in 2022 by a team of Oxford University scientists, Fractile~\cite{fractile} aims to transform the world by enabling large language models' (LLM) inference at speeds up to 100 times faster than Nvidia's most recent H100 GPUs. This increase in performance primarily arises from in-memory computations. However, the details of the technology, both hardware and software, as well as the company's funding particulars, remain undisclosed. 

\subsection{Untether AI}
\label{subsec:untether}
Founded in 2018~\cite{untether}, Untether's main design integrates RISC-V cores on the SRAM chips for processing AI workloads. Their latest product, the tsunAImi accelerator card provides a phenomenal 2 POPS of compute power, twice the amount of any available product. This compute power translates into over 80,000 frames per second of ResNet-50 throughput, three times the throughput of any product on the market.  Untether AI provides an \underline{automated SDK} for its products. The SDK takes a network model implemented in common machine learning frameworks like TensorFlow and PyTorch and lowers it into the kernel code that runs on these RISC-V processors. It automatically takes care of low-level optimizations, providing extensive visualization, a cycle-accurate simulator, and an easily adoptable runtime API.

\vspace{0.1cm}\noindent\textbf{Funding:} Untether's investors include CPPIB, GM Ventures, Intel Capital, Radical Ventures, and Tracker Capital.  
\subsection{UPMEM Technology}
\label{subsec:upmem-commercial}
Founded in 2015, UPMEM is a tech company offering programmable CNM systems for data-intensive applications. See more details on the architecture and programmability in Section~\ref{sss:upmem}.

\vspace{0.1cm}\noindent\textbf{Funding:} The company is funded by Western Digital, Partech, and super nova invest.

\subsection{Summary}
\label{subsec:landscape-summary}
\begin{figure*}[tbh]
\centering
\includegraphics[scale=5.5]{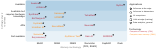}
\caption{A landscape of CIM and CNM companies, highlighting their technologies, target applications and software stack readiness.} 
\label{fig:landscape}
\end{figure*}

Table~\ref{fig:landscape-table} and Figure~\ref{fig:landscape} summarize the discussion in this section and provides a landscape of CIM, CNM companies, their products, technologies, and funding status. Please note that this compilation is not exhaustive; it includes only companies known to us and those that, based on our understanding, fall within the CIM and CNM categories. As Figure~\ref{fig:landscape} clearly illustrates, the current landscape is predominantly characterized by conventional technologies, with a notable absence of a comprehensive software ecosystem.

\begin{table}
\scriptsize
    \begin{center}
    \begin{tabular}{p{0.09\linewidth}p{0.13\linewidth}p{0.18\linewidth}p{0.13\linewidth}
    p{0.17\linewidth}p{0.17\linewidth}}
    \toprule
        \textbf{Company} & \textbf{Use-Case} & \textbf{Technology} & \textbf{Solution} & \textbf{Programmability} & \textbf{Funding (Mio. \$) \hspace{1cm}(PitchBook)}\\ 
        \midrule
        Axelera & AI on the Edge  & SRAM (digital MAC)& Hardware-SoC & SDK provided & 63.72 (Early stage VC) \\ 
        d-Matrix & AI inference in datacenters & SRAM (digital MAC) & Chiplets  & Open-source Framework(s) & 161.3 (Early stage VC) \\ 
             Synthara & AI, DSP, Linear algebra& SRAM (dot product) & 
        Accelerator & Compiler available & 3.33 (Grant) \\
    
        Mythic & AI on the Edge & Flash (Analog computing)& Accelerator, processor &
        Software stack  (does rewriting, opt, mapping) & 177.41 (Later stage VC) \\ 
    
        Surecore & AI on the edge & SRAM (CNM) & Chip & No details & 11.16 (Later stage VC)\\
    
        SEMRON & AI on the edge & Memcapacitor & 3D-Chip (planned) & No details & 1.63 (Seed round)\\ 
    
        Untether AI & AI everywhere & SRAM+RISC-V (CNM) & Chips, accelerator & SDK and simulator (Toolkit) & 
        153.52 (Early stage VC)\\ 
        
        Syntiant & AI on the edge & SRAM+ARM MCUs (CNM) & Processor & Available & 121.43 (Later stage VC) \\
    
        Neuroblade & Analytics & DRAM+RISC cores (CNM) & Processor & Set of APIs & 110.43 (Debt - General)\\ 
    
        Rain AI & LLMs on the edge (Training)&  Memristors & Processors & NA & 64.04 (Later stage VC)\\ 
    
        TetraMem & Edge applications & Memristors & Processors, software stack& HDK, SDK & NA \\ 
    
        Gyrfalcon Tech & AI on the edge & CNM (MACs with MRAM) & Chip & NA & 68.0 (Debt - PPP) \\ 
    
        UPMEM & General-purpose & DRAM+RISC cores & System & APIs & 15.5 (Later stage VC)\\ 
    
    
        EnCharge AI & AI inference & SRAM (CIM) + SIMD (CNM) & Chip & 
        Software available & 21.7 (Angel - individ) \\ 
    
        Re(conveive) & AI inference & SRAM (analog CIM) & Chip & NA & NA \\ 
    
        Fractile & LLMs inference & NA & Chip & NA & NA \\
\bottomrule
\end{tabular}
    \caption{CIM/CNM companies, with their products, technologies, and funding status.} 
    \label{fig:landscape-table}
\end{center}

\end{table}

\subsection{Open challenges}
\label{subsec:challenges}
\ac{cim} and \ac{cnm} systems have already entered the market, yet a series of open challenges are expected to become more pronounced as time progresses. It will take years to understand how these units will harmonize within the overall system architecture and determine their optimal utilization. 
In the following, we briefly discuss the important categories.

~\\\noindent\textbf{Materials:} During the era of Moore's law in computing, the primary focus was on refining transistors to be smaller, faster, and more energy-efficient. The selection of materials was confined to only those compatible with manufacturing processes. However, the limitations of these materials to scale further are now exposed. As a result, new materials have emerged and further research is needed to investigate novel materials (to enable further transistor scaling: hopes with carbon nanotube, and novel memory devices).
~\\\noindent\textbf{Devices:} Mainstream computing has largely relied on digital logic and binary storage. Nonetheless, the emerging wave of computing architectures, particularly \ac{cim} requires novel multi-state devices allowing both analog and digital operations. Existing devices, memristors in particular, do offer such properties but have reliability and other associated challenges. 
~\\\noindent\textbf{Integration:} We have seen various architectures based on various technologies. As is evident, there is no on-technology-fits-all solution. Eventually, \ac{cim} modules based on different technologies need to be integrated into the same to get the best out of all these technologies. This poses integration challenges that have received little to no attention. 
~\\\noindent\textbf{Processing systems:} These novel architectures require new tools, algorithms, cost models, and software solutions. All of them are crucial to understanding these architectures, enabling their design space exploration, and making them accessible to a larger audience.

While every challenge holds significance and demands attention, programmability and user-friendliness are the most important ones from the user's standpoint. Following is an excerpt from Facebook's recent article on their inference accelerator that highlights the same.

\emph{``We’ve investigated applying processing-in-memory (PIM) to our workloads and determined there are several challenges to using these approaches. Perhaps the biggest challenge of PIM is its programmability''.}

In response to the challenges associated with programmability, we have ourselves been working on high-level programming and compilation frameworks for CNM and CIM systems~\cite{khan_tcad21, khan2022cinm, farzaneh2023c4cam,delima_date24}. 
We have developed reusable abstractions and demonstrated compilation flows for CIM systems with memristive crossbars, CAMs, CIM-logic modules, and for CNM systems like UPMEM and Samsung CNM.
However, much more cross-layer work is needed to improve automation~\cite{niemier_date24}, in particular for heterogeneous systems integrating several paradigms and technologies. 

%% file: contents/conclusions.tex
This paper overviews the landscape of compute-near-memory (CNM) and compute-in-memory (CIM) paradigms. It starts with an explanation of the Von Neumann bottleneck, the necessity of novel CIM/CNM paradigms, and the key terminology used in the related literature. It offers a comprehensive background on major memory technologies and emphasizes the importance of heterogeneous systems. The paper overviews prominent CIM and CNM designs from both academia and industry. In contrast to other studies in the literature that focus on either application domains or memory technologies, this paper concentrates on designs that have either successfully transitioned into product offerings or have reached a stage where commercialization is a feasible prospect. We explain prevalent CNM architectures, including microarchitectural details, associated technologies, software frameworks, and the results achieved (usually measured as throughput). Subsequently, we survey the landscape of CIM systems, explaining prevailing CIM designs that use prominent technologies such as SRAM, DRAM, MRAM, RRAM, PCM, and FeFET.  We overview CIM chips from industrial giants (research centers), spanning from earlier designs like ISAAC and PUMA by Hewlett Packard Enterprise to the most cutting-edge chips from IBM, Samsung, TSMC, Intel, Meta (Facebook), Bosch, Frauenhofer, and GlobalFoundries. Current trends in industrial research show that while conventional SRAM and DRAM technologies are ready to be leveraged in CIM/CNM systems, emerging technologies like PCM, RRAM, MRAM, and FeFETs are also poised to make partial inroads, particularly for selected operations, such as dot products and pattern matching.

Finally, we describe the landscape of CIM and CNM start-ups, highlighting the emergence of numerous new companies in recent years that have introduced innovative solutions to cater to the thriving demands of AI and other data-intensive application domains. These companies are targeting a diverse range of market segments, spanning from power-efficient edge applications (AI at the edge) to high-performance data center servers (e.g., for AI training), and many have successfully secured substantial funding (hundreds of millions) in their initial funding rounds.
The paper shows that SRAM technology currently dominates this landscape. However, with active research and breakthroughs in emerging NVMs (demonstrated by recent industrial chips), it is anticipated that NVMs will play a more prominent role in these paradigms in the near future.

The paper highlights that CIM and CNM technologies (i) harbor significant potential to outperform conventional systems, and (ii) have already made inroads into the market. However, their true potential remains untapped. This is attributed to a number of challenges, including the lack of accurate design space exploration tools, programming frameworks, and a comprehensive software ecosystem in general, and cost and performance models that can be leveraged to guide static and runtime optimizations for these systems.

CNM and CIM computing is an extremely active field. 
We believe that we have captured a representative snapshot of this field, early in year 2024, and remain excited about how technologies, devices, architectures and tools will continue to develop moving forward.

%% file: acronyms.tex
\newpage
\onecolumn
\thispagestyle{empty}
\section*{Acronyms}

\renewenvironment{description} 
{\list{}{\labelwidth0pt\itemindent-\leftmargin
    \parsep0pt\itemsep0pt\let\makelabel\descriptionlabel}}
               {\endlist}

\begin{acronym}
  \acro{cim}[CIM]{\emph{compute-in-memory}}
  \acro{imc}[IMC]{\emph{in-memory-computing}}
  \acro{imp}[IMP]{\emph{in-memory-processing}}
  \acro{lim}[LIM]{\emph{logic-in-memory}}
  \acro{pim}[PIM]{\emph{processing-in-memory}}
  \acro{pum}[PUM]{\emph{processing-using-memory}}
\end{acronym}

\begin{acronym}
  \acro{cnm}[CNM]{\emph{compute-near-memory}}
  \acro{nmc}[NMC]{\emph{near-memory-computing}}
  \acro{pnm}[PNM]{\emph{processing-near-memory}}
  \acro{nmp}[NMP]{\emph{near-memory-processing}}
\end{acronym}

\begin{acronym}
\acro{adc}[ADC]{\emph{analog-to-digital converter}}
\acro{ai}[AI]{\emph{artificial intelligence}}
\acro{apu}[APU]{\emph{accelerated processing unit}} 
\acro{bl}[BL]{\emph{bitline}}
\acro{blb}[BLB]{\emph{bitline bar}}
\acro{cim-a}[CIM-A]{CIM-array}
\acro{cim-p}[CIM-P]{CIM-peripheral}
\acro{cam}[CAM]{\emph{content-addressable-memory}}
\acro{cnn}[CNN]{\emph{convolutional neural network}}
\acro{cmos}[CMOS]{\emph{complementary metal–oxide–\-se\-miconductor}} 
\acro{com}[COM]{\emph{compute-outside-memory}}
\acro{cpu}[CPU]{\emph{central processing unit}}
\acro{dac}[DAC]{{digital-to-analog converter}}
\acro{dnn}[DNN]{\emph{deep neural network}}
\acro{dpu}[DPU]{\emph{data processing unit}}
\acro{dram}[DRAM]{\emph{dynamic random-access memory}}
\acro{fefet}[FeFET]{\emph{ferroelectric field-effect transistor}}
\acro{fpu}[FPU]{\emph{floating-point unit}}
\acro{hbm}[HBM]{\emph{high bandwidth memory}}
\acro{hmc}[HMC]{\emph{hybrid memory cube}}
\acro{isa}[ISA]{\emph{instruction set architecture}}
\acro{lut}[LUT]{\emph{look-up table}}
\acro{mac}[MAC]{\emph{multiply-accumulate}}
\acro{ml}[ML]{\emph{machine learning}}
\acro{mlc}[MLC]{\emph{multi-level cell}}
\acro{mos}[MOS]{\emph{metal-oxide-semiconductor}}
\acro{mram}[MRAM]{\emph{magnetic RAM}}
\acro{mtj}[MTJ]{\emph{magnetic tunnel junction}}
\acro{mpsoc}[MPSoC]{\emph{multiprocessor system-on-chip}}
\acro{nn}[NN]{\emph{neural-network}}
\acro{nvm}[NVM]{\emph{non-volatile memory}}
\acro{omp}[OMP]{\emph{outside memory processing}}
\acro{pcm}[PCM]{\emph{phase change memory}}
\acro{pc}[PC]{\emph{program counter}}
\acro{pu}[PU]{\emph{processing unit}}
\acro{rfu}[RFU]{\emph{eserved-for-
future-use}}
\acro{rram}[RRAM]{\emph{resistive RAM}}
\acro{rtm}[RTM]{\emph{racetrack memory}}
\acro{sdk}[SDK]{\emph{software development kit}}
\acro{simd}[SIMD]{\emph{single-instruction multiple-data}}
\acro{slc}[SLC]{\emph{single-level cell}}
\acro{soc}[SoC]{\emph{system-on-chip}}
\acro{sram}[SRAM]{\emph{static random-access memory}}
\acro{sot}[SOT]{\emph{Spin-orbit-torque}}
\acro{stt}[STT]{\emph{spin-transfer-torque}}
\acro{stt-mram}[STT-MRAM]{\emph{spin-transfer-torque magnetic memory}}
\acro{tmr}[TMR]{\emph{tunnel magnetoresistance ratio}}
\acro{tra}[TRA]{\emph{triple row activation}}
\acro{tsvs}[TSVs]{\emph{through-silicon vias}}
\acro{wl}[WL]{\emph{word line}}
\acro{cagr}[CAGR]{\emph{compound annual growth rate}} 
\acro{mvm}[MVM]{\emph{matrix-vector multiplication}} 
\acro{sa}[SA]{\emph{sense amplifier}} 
\acro{llm}[LLM]{\emph{large language models}}

\end{acronym}